%% file: icde.tex
\documentclass[conference]{IEEEtran}
\IEEEoverridecommandlockouts
\makeatother
\newtheorem{example}{Example}
\usepackage{cite}
\usepackage{makecell}
\usepackage{soul}
\usepackage{amsmath,amssymb,amsfonts}

\usepackage{algorithmic}
\usepackage[ruled,vlined, linesnumbered]{algorithm2e}
\usepackage{graphicx}
\usepackage{textcomp}
\usepackage{xcolor}
\usepackage{enumitem}
\def\BibTeX{{\rm B\kern-.05em{\sc i\kern-.025em b}\kern-.08em
    T\kern-.1667em\lower.7ex\hbox{E}\kern-.125emX}}
\usepackage{enumitem}
\usepackage{pifont}

\usepackage{cleveref}
\crefname{section}{§}{§§}
\Crefname{section}{§}{§§}

\newcommand{\shadd}[1]{\textcolor{black}{#1}}

\definecolor{LightCyan}{rgb}{0.88,1,1}
\definecolor{LightRed}{rgb}{1,0.88,1}
\definecolor{LightYellow}{rgb}{1,1,0.88}
\definecolor{LightGray}{gray}{0.8}
\long\def\comment#1{}
\input{custom_sytles}

\input{command}
\begin{document}

\title{All-in-One: Heterogeneous Interaction Modeling for Cold-Start Rating Prediction}

\author{
\IEEEauthorblockN{Shuheng Fang\IEEEauthorrefmark{1},
Kangfei Zhao\thanks{\IEEEauthorrefmark{2} Corresponding author.}\IEEEauthorrefmark{2},
Yu Rong\IEEEauthorrefmark{3},
Zhixun Li\IEEEauthorrefmark{1} and
Jeffrey Xu Yu\IEEEauthorrefmark{1}}
\IEEEauthorblockA{\IEEEauthorrefmark{1}\textit{The Chinese University of Hong Kong},
\{shfang, yu\}@se.cuhk.edu.hk, lizhixun1217@gmail.com}
\IEEEauthorblockA{\IEEEauthorrefmark{2}\textit{Beijing Institute of Technology},
zkf1105@gmail.com}
\IEEEauthorblockA{\IEEEauthorrefmark{3}\textit{Alibaba DAMO Academy},
yu.rong@hotmail.com}}

\comment{
\author{
\IEEEauthorblockN{Shuheng Fang}
\IEEEauthorblockA{\textit{The Chinese University of Hong Kong} \\
shfang@se.cuhk.edu.hk}
\and
\IEEEauthorblockN{Kangfei Zhao}
\IEEEauthorblockA{\textit{Beijing Institute of Technology} \\
zkf1105@gmail.com}
\and
\IEEEauthorblockN{Yu Rong}
\IEEEauthorblockA{\textit{Tencent AI Lab} \\
yu.rong@hotmail.com}
\and
\IEEEauthorblockN{Zhixun Li}
\IEEEauthorblockA{\textit{The Chinese University of Hong Kong} \\
lizhixun1217@gmail.com}
\and
\IEEEauthorblockN{Jeffrey Xu Yu}
\IEEEauthorblockA{\textit{The Chinese University of Hong Kong} \\
yu@se.cuhk.edu.hk}
}}

\maketitle
\thispagestyle{plain}

\begin{abstract}

Cold-start rating prediction is a fundamental problem in recommender systems that has been extensively studied. Many methods have been proposed that exploit explicit relations among existing data, such as collaborative filtering, social recommendations and heterogeneous information network, to alleviate the data insufficiency issue for cold-start users and items. However, the explicit relations constructed based on data between different entities may be unreliable and irrelevant, which limits the performance ceiling of a specific recommendation task. Motivated by this, in this paper, we propose a flexible framework dubbed heterogeneous interaction rating network (\HIRE). \HIRE does not solely rely on pre-defined interaction patterns or a manually constructed heterogeneous information network. Instead, we devise a Heterogeneous Interaction Module (\HIM) to jointly model heterogeneous interactions and directly infer the important interactions via the observed data. In the experiments, we evaluate our framework under 3 cold-start settings on 3 real-world datasets. The experimental results show that \HIRE outperforms other baselines by a large margin. Furthermore, we visualize the inferred interactions of \HIRE to reveal the intuition behind our framework.


\end{abstract}


\section{Introduction}

Rating prediction is a fundamental task in recommender systems for estimating  users' preference scores on items, which has widespread applications~\cite{lyu2022memorize,guo2022miss,DBLP:conf/icde/QianL0023,DBLP:conf/icde/LiWLS23,cao2022cross} across e-commerce, online education, and entertainment platforms.  In cold-start scenarios, where new users/items have limited information for prediction, accurate rating prediction is crucial for effective user retention (in the case of cold users) and item promotion  (in the case of cold items). 
The scarcity of ratings and interactions in cold-start scenarios poses  significant challenges, as current recommendation models heavily rely on  observed ratings and interactions to make predictions. To improve prediction accuracy, current solutions incorporates explicit additional associations and side information to enrich the features of cold users and items. These approaches can be generally categorized into three lines of research.

\begin{figure}[t]
\centering
	\includegraphics[width=0.4\textwidth]{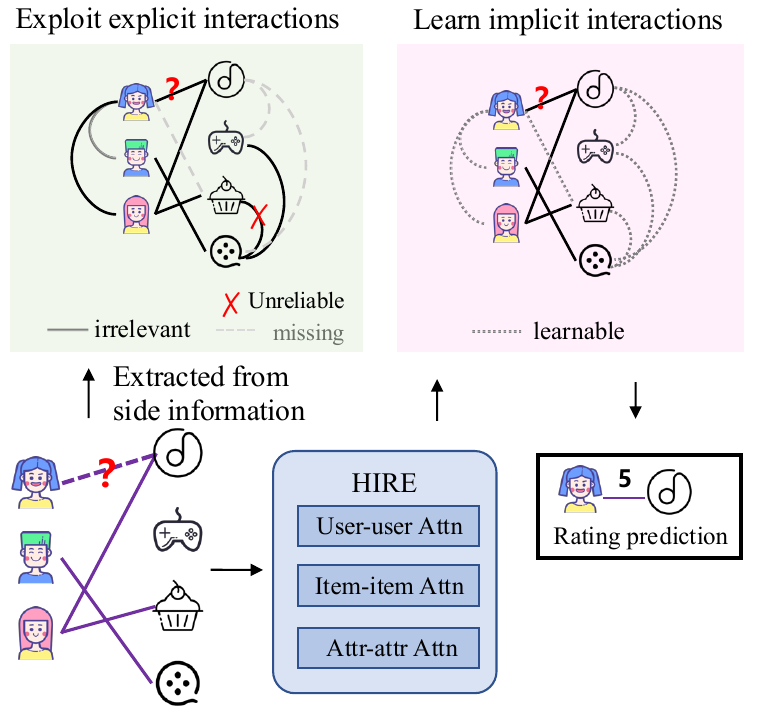}
	\caption{ {An illustrative example: 
    to predict the first girl's preference for music, relation between the girl and a boy seems to be irrelevant. The erroneous classification for cake (food) and movie (entertainment) is unreliable for rating prediction. 
    \HIRE directly models heterogeneous interactions in a data-driven way.}
 }
  \label{fig:motivation}
\end{figure}

\shadd{Collaborative filtering (CF)~\cite{CF1,CF2,NCF,CF3,CF4,CF5,rong2014monte}, the most traditional approach, assumes similar users and items exhibit similar ratings and exploits the user-item interactions to make the rating prediction. However, CF models are limited to process single-type interactions and fail to generalize well to cold users/items with scarce rating interactions.
Social recommendation approaches~\cite{social,wu2019neural, fan2019graph,trustwalker,exploiting,sorec,jamali2010matrix} extend the scope of interactions and introduce social connections into the recommendation framework. These approaches are on the basis that users' preferences are highly influenced by their social communities and opinion leaders within social networks. However, the effectiveness of such approaches is constrained by the availability and reliability of social relations.
Furthermore, recent research has explored the integration of complex interactions through Heterogeneous Information Networks (HINs), which incorporate entities beyond users and items, such as tags and points of interest (POIs). While HINs promise richer semantic information and enhance the representations of users and items, HIN-based approaches rely on pre-defined graph structures and are therefore vulnerable to the noise and mistakes introduced during graph construction. They also suffer from the issue of irrelevance, similar to social recommendation approaches. Fig.~\ref{fig:motivation} illustrates the explicit interactions in the system and the challenge of effectively exploiting them.
}

\shadd{
To address the issue of cold users/items, recent studies \cite{TaNP, MAMO, MeLU} employ meta learning to accommodate cold-start scenarios. MeLU \cite{MeLU} utilizes model-agnostic meta-learning (MAML) \cite{MAML} to develop a personalized model for each user based on its preferences for a limited number of items. TaNP \cite{TaNP} develops a task-adaptive mechanism for learning the relevance of different users. MAMO \cite{MAMO} devises user preference memory to capture shared  preferences. 
Despite these advances, these approaches  capture only explicit interactions between users rating the same items, missing implicit relationships. 
In addition, their complicated optimization or memorization strategies in meta-learning-based recommendation will incur the barrier of implementations. 
}

To summarize, the homogeneity, unreliability and irrelevancy introduced by the explicit interactions present significant obstacles for existing recommendation approaches, especially for the cold-start scenarios. This motivates us to reconsider \emph{how a rating model can automatically infer heterogeneous interactions and how rating prediction can be supported in an end-to-end fashion.}

In this paper, we propose a deep learning framework named \underline{H}eterogeneous \underline{I}nteraction \underline{R}ating N\underline{e}twork (HIRE) for rating prediction in cold-start scenarios. 
{Distinguished from existing approaches we discussed, HIRE does not rely on side information in the data, but models the heterogeneous interactions for rating prediction in a fully data-driven way.
Specifically, the core component of our HIRE model is the Heterogeneous Interaction Module (HIM), which jointly learns heterogeneous interactions from a prediction context of the cold users/items. HIM consists of three multi-head self-attention (MHSA) layers, the key building block of foundation models that exhibits powerful modeling capability of interactions in NLP tasks. However, in contrast to the usage of textual sequence, we leverage this interaction modeling capability of MHSA to capture the correlations between the users, items and attributes in the prediction context. The inherent metric learning idea of MHSA is beneficial for cold-start scenarios where the target users/items have only a few interactions.
For ratings of user-item pairs to be predicted, to construct a prediction context, we design a neighborhood-based sampling strategy to select relevant users and items from the user-item rating bipartite graph. This provides an initial prior that is informative for predicting the ratings of cold users and items. } 
The model is trained by learning the underlying heterogeneous interactions from multiple sampled prediction contexts to predict masked ratings. 
Our accurate prediction results demonstrate the promise of holistically modeling heterogeneous interactions using attention layers for cold-start recommendation tasks.
The contributions of this paper are summarized as follows: 
\begin{itemize}[leftmargin=*]
\item We propose a novel learning framework, \HIRE, for the cold-start rating prediction in recommendation systems. 
Instead of explicitly exploiting external interaction information, \HIRE learns the interactions from heterogeneous sources in a holistic and data-driven fashion, and can deal with 3 typical cold-start scenarios, i.e., user cold-start, item cold-start and both user and item cold-start.
\item We design a Heterogeneous Interaction Module (\HIM), powered by multi-head self-attention. 
\HIM learns the interaction in the perspective of users, items, and the multi-category attributes from a prediction context, which are composed of a set of users and items. 
We devise a sampling-based, simple yet effective context construction strategy and prove that \HIM is permutation equivariant to the order of the users or items in the prediction context. 
\item \shadd{We conduct substantial experimental studies on 3 real-world datasets in the 3 cold scenarios to verify the effectiveness of \HIRE. Compared with 12 baselines, our \HIRE outperforms others with higher Precision, NDCG and MAP by 0.21, 0.29 and 0.22 on average.} 
We conduct case studies, and show that \HIM provides potential interpretive ability for rating prediction.
\end{itemize}

\input{related}
\input{preliminary}

\input{methodology}

\input{experiment}

\section{Conclusion}
\label{sec:conclusion}
In this paper, we propose a fully data-driven recommender system \HIRE that can model heterogeneous interactions for cold-start rating prediction. 
For target cold user/items, we adopt a neighborhood-based sampling strategy to sample prediction context from the user-item bipartite rating graph, and explore reliable interactions in the context. 
Specifically, we devise a Heterogeneous Interaction Module to jointly model the interactions in three levels, i.e., users, items and attributes, respectively. Comprehensive experiments on 3 real-world datasets validate the effectiveness
of our model. {HIRE outperforms baselines
with higher precision, NDCG and MAP by 0.21, 0.29 and 0.22 on average.} {The source code is publicly available at \url{https://github.com/FangShuheng/HIRE}.}


\section*{Acknowledgments}
This work was supported by the Research Grants Council of Hong Kong, China, No. 14203618, No. 14202919 and No. 14205520.
Kangfei Zhao is supported by National Key Research and Development Plan, No. 2023YFF0725101.

\bibliographystyle{IEEEtran}
\bibliography{ref}

\end{document}

%% file: custom_sytles.tex
\usepackage{graphicx}
\usepackage[linesnumbered,ruled]{algorithm2e}
\usepackage{lipsum}
\usepackage{amsmath}
\usepackage{physics}
\usepackage{subfigure}
\usepackage{tabularx}
\usepackage{color}
\usepackage{colortbl}
\usepackage{float}
\usepackage{listings}
\usepackage{multirow}
\usepackage[normalem]{ulem}
 \usepackage{longtable}
\usepackage{enumitem}
\usepackage{multirow}
\usepackage{booktabs}
\usepackage{capt-of}
\usepackage{pifont}
\usepackage{ulem}
\usepackage{soul}
\usepackage{cancel}
\usepackage{bm}
\usepackage{url}
\usepackage{caption}
\usepackage{tikz}

\def\BibTeX{{\rm B\kern-.05em{\sc i\kern-.025em b}\kern-.08em
    T\kern-.1667em\lower.7ex\hbox{E}\kern-.125emX}}

 \graphicspath{{./Graph/}, {./Fig/}, {./Legend/}}

\newcounter{definition}[section]
\renewcommand{\thedefinition}{\nthesection.\arabic{definition}}

\newcounter{theorem}[section]
\renewcommand{\thetheorem}{\nthesection.\arabic{theorem}}

\newcounter{lemma}[section]
\renewcommand{\thelemma}{\nthesection.\arabic{lemma}}
        
\newcounter{proposition}[section]
\renewcommand{\theproposition}{\nthesection.\arabic{proposition}}

\newcounter{remark}[section]
\renewcommand{\theremark}{\nthesection.\arabic{remark}}

\newcounter{property}[section]
\renewcommand{\theproperty}{\nthesection.\arabic{property}}
\newenvironment{property}{
     \refstepcounter{property}
     {\vspace{1ex} \noindent\bf  Property  \theproperty.}}{
     \vspace{1ex}} 

\newcommand{\myproof}{\noindent{\bf Proof: }}
\newcommand{\nthesection}{\arabic{section}}

\newcommand{\eop}{\hspace*{\fill}\mbox{$\Box$}\vspace*{1ex}}

\newcommand{\stitle}[1]{\vspace{1ex} \noindent{\bf #1}}

\newcommand{\green}[1]{\textcolor{green}{}}

\newcommand{\kw}[1]{{\ensuremath {\mathsf{#1}}}\xspace}

%% file: command.tex
\newcommand{\etitle}[1]{\vspace{1ex} \noindent{\underline{\em #1}}}

\newcommand{\loss}{\mathcal{L}}

\newcommand{\ABU}{{MBU}\xspace}
\newcommand{\ABI}{{MBI}\xspace}
\newcommand{\ABF}{{MBA}\xspace}

\newcommand{\Popularity}{{{Popularity}}\xspace}
\newcommand{\transgnn}{{{TransGNN}}\xspace}
\newcommand{\TANP}{{{TaNP}}\xspace}
\newcommand{\MELU}{{{MeLU}}\xspace}
\newcommand{\MAMO}{{{MAMO}}\xspace}
\newcommand{\MetaHIN}{{{MetaHIN}}\xspace}
\newcommand{\NEUMF}{{{NeuMF}}\xspace}
\newcommand{\DEEPFM}{{{DeepFM}}\xspace}
\newcommand{\WIDE}{{{Wide\&Deep}}\xspace}

\newcommand{\AFN}{{{AFN}}\xspace}
\newcommand{\CT}{{{HIRE}}\xspace}
\newcommand{\HIRE}{{{HIRE}}\xspace}
\newcommand{\HIM}{{{HIM}}\xspace}
\newcommand{\GraphRec}{{{GraphRec}}\xspace}
\newcommand{\GraphHINGE}{{{GraphHINGE}}\xspace}

\newcommand{\PRE}{\kw{Pre.}}
\newcommand{\MAP}{\kw{MAP.}}
\newcommand{\NDCG}{\kw{NDCG.}}

\newcommand{\UIC}{{U\&I~C}\xspace}
\newcommand{\UC}{{UC}\xspace}
\newcommand{\IC}{{IC}\xspace}

\newcommand{\Movielens} {\kw{MovieLens}-\kw{1M}}
\newcommand{\Bookcrossing} {\kw{Bookcrossing}}
\newcommand{\MovieTweetings} {\kw{MovieTweetings}}
\newcommand{\Douban} {\kw{Douban}}

\newcommand{\UserSet}{{\mathcal{U}}}
\newcommand{\ItemSet}{{\mathcal{I}}}
\newcommand{\RateSet}{{\mathcal{R}}}

\newcommand{\Real}{\mathbb{R}}
\newcommand{\BigO}{{\mathcal{O}}}

%% file: related.tex
\section{Related Work}
\label{sec:related}


\stitle{CF-based approaches.} Collaborative Filtering (CF) is a classical approach in recommendation systems. In recent years,  CF is enhanced by deep learning, where neural networks enable feature fusion from auxiliary information, e.g., user profiles and item descriptions. 
NCF \cite{NCF} leverages multi-layer perceptron to capture non-linear feature interaction between user and item and can generalize matrix factorization. 
Wide\&Deep \cite{WIDE} incorporates wide linear models and deep neural networks to fulfill memorization of sparse feature interaction and generalization to unseen user and item features. 
DeepFM \cite{deepFM} integrates factorization machine and neural networks for jointly modeling the low-order and high-order user-item feature interactions.
AFN \cite{AFN} introduces a logarithmic transformation layer with neural networks to model arbitrary-order feature interactions for rating prediction. 
The CF-based approaches only model the interactions of user and item features, and are not specifically tailored for cold-start recommendation. 

\stitle{Social recommendation approaches.}
Social recommendation relies on additional relationships~\cite{fang2023community,fang2024inductive,zhang2024adaptivecoordinatorspromptsheterogeneous} from social media to enhance feature interactions.
TrustWalker \cite{trustwalker} proposes a random walk model to combine the interactions in trust networks among users with item-based collaborative filtering. 
MCCP~\cite{rong2014monte} designs the random walk process on user-item bipartite graph to simulate the preference propagation to alleviate the data sparsity problem for cold-start users.
LOCABLE \cite{exploiting} exploits social contents from both local user friendship and global ranking of user reputation, i.e., PageRank.
SoRec \cite{sorec} jointly factorizes user-item rating matrix and user-user social relation matrix.
GraphRec \cite{fan2019graph} is a GNN framework that jointly models the social influence in a user-user graph and the rating opinions in a user-item graph.
To better learn user and item embeddings for recommendation, DiffNet \cite{wu2019neural} simulates the social influence propagation in a social network by an influence diffusion neural network model.
Leveraging social information alleviates data scarcity and sparsity in cold-start scenarios. However, these approaches rely on explicit and external social information, which can be incomplete and inaccurate.

\comment{
\begin{table}[t]
\small
\centering
\caption{Frequently-used Notations}
\resizebox{0.45\textwidth}{!}{
\label{tab:notation}
\begin{tabular}{c|c}
\toprule
Notation & Description                 \\\midrule
$\mathcal{U}$/$\mathcal{I}$         & user/item set in the recommending system \\
$\mathcal{R}$ & observed rating in the system \\
 $u$/$i$ & a user / an item \\
 $m$/$n$ & number of users/items in each context \\
 $\bm{e}_u$/$\bm{e}_i$ & one-hot embedding of the attributes of $u$/$i$ \\
 $h$ & number of attributes of user and item\\
 $f$ & embedding dimension of each attribute\\
 $e$ & embedding dimension for a pair of user and item\\
$f_U^k$/$f_I^k$ & linear transformation for the $k$-th attribute of user/item\\
$f_R$ & rating linear transformation\\
 $\bm{H}$ & initial embedding of a prediction context \\
 $\hat{R}$ & predicted rating matrix \\
 \bottomrule
\end{tabular}}
\end{table}
}

\stitle{HIN-based approaches.} Similar to social recommendation approaches, HIN-based approaches further utilize heterogeneous interactions from HINs.
In early studies, meta-path or meta-graph based user and item similarity matrices are utilized for matrix factorization~\cite{heterec,DBLP:conf/kdd/ZhaoYLSL17} and collaborative filtering~\cite{luo2014hete}. 
GraphHINGE~\cite{jin2020efficient} designs a neighborhood-based interaction model to enhance the user and item representations, where the neighbors are selected by  meta-paths~\cite{sun2011pathsim}.
Similarly, NI-CTR \cite{neighbor} leverages neighborhood interactions in 4 types of HIN to assist the CTR prediction.
To address cold-start problem, 
MetaHIN \cite{metahin} exploits the rich semantics in HINs at the data level and meta-learning at the model level.
HIN-based approaches need to collect and construct HINs with manually defined heterogeneous patterns, i.e., meta-paths or meta-graphs for any specific dataset. 
In contrast, our approach HIRE learns the heterogeneous interactions in a generic and data-driven way.


\stitle{Cold-start recommendation.} 
Meta-learning~\cite{snell2017prototypical,ravi2016optimization,graves2014neural,MAML} aims to learn a meta model that can rapidly adapt to new tasks with a few training samples.
Recently, meta-learning is used for cold-start recommendation where predication for one user is treated as a task. 
MeLU \cite{MeLU} adopts model-agnostic meta learning (MAML) \cite{MAML} to learn a personalized model for each user given her preference on a few items.
MAMO \cite{MAMO} also adopts MAML to learn a global parameter initialization for all users. In addition, a feature-specific memory module and a task-specific memory module are used to guide parameter personalization and fast prediction of user preference.
TaNP \cite{TaNP} learns a Neural Process \cite{garnelo2018neural}, an encoder-decoder model architecture for meta-learning, where the decoder is equipped with a task-adaptive mechanism for task-specific adaptation.
%
These proposed meta model only explicitly model the interaction of user-item pairs, where the potential interactions among users are derived from model adaptation. \comment{Our approach can be generalized to a meta model when a prediction context is regarded as a task. The difference is that in predication contexts, we explicitly model extra interactions among multiple users and items.} 

{
\shadd{Recently, Transformer-based methods \cite{jiang2023adamct,kang2019recommender,kang2018self} are proposed for capturing user-item interactions adaptively and globally. TransGNN \cite{zhang2024transgnn} integrates Transformer and
GNN layers to enhance the Transformer’s performance.}
Following the scaling law of Transformers, Large language models (LLMs) have presented strong zero-shot generalization ability in cold-start recommendation~\cite{wu2023survey, wang2024large, sanner2023large}. 
However, LLMs for recommendation require user and item possessing rich semantic descriptions as extra knowledge. Furthermore, the performance of LLMs degenerates in numerical prediction tasks. 
}

\comment{
\subsection{Collaborative Filtering for Cold-strat Recommendation}
\label{subsec:CSrecommendation}
CF-based methods are widely used to explore potential interactions. 
NCF \cite{NCF} exploits DNN to capture non-linear feature interaction between user and item. 
Wide\&Deep \cite{WIDE}, combining wide linear models and deep neural networks to memorize sparse feature interactions and generalize to previously unseen feature interactions through low-dimensional embeddings.
DeepFM \cite{deepFM}, similar to Wide\&Deep, but has a shared input to its “wide” and “deep” parts.
AFN \cite{AFN} uses logarithmic transformation layer to model arbitrary-order feature interactions for rating prediction.  

Recently, meta-learning framework for cold-start recommendation has achieved competitive performance. These works introduce additional modules to augment existing data by modeling complex interactions. 
TaNP \cite{TaNP} maps observed ratings of each user to a predictive distribution and introduce a task-adaptive mechanism. It enables their model to learn relevance of different users and helpful to estimate user preference. 
MeLU \cite{MeLU} adopts the framework of MAML \cite{MAML} in recommender system and provide personalized model to each user with their item-consumption history.
MAMO \cite{MAMO} also adopts MAML for parameter initialization and they design two memory matrices to store task-specific memories and feature-specific memories. The task-specific memories can guide model predicting user preference.
They consider additional implicit interactions among users compared to traditional CF-based methods. However, it is not enough. More comprehensive interactions should be utilized to help rating prediction.

\subsection{Social Recommendation for Cold-start}
Social recommendation relies on additional social relations to enhance existing interactions. 
TrustWalker \cite{trustwalker} proposes a random walk model to explore direct and indirect ratings of friends for the target item as well as similar items. 
LOCABLE \cite{exploiting} peform co-factorization to capture local and global social relations.
SoRec \cite{sorec} jointly factorize user-item rating matrix and user-user social relation matrix.
GraphRec \cite{fan2019graph} presents a graph neural network framework to model two graphs and heterogeneous strengths for social recommendation.
DiffNet \cite{wu2019neural} design layer-wise influence diffusion structure for users and models users' latent embeddings evolution.
Social recommendation can also address cold-start scenarios.
\cite{jamali2010matrix} make recommendations for users in a large social network since these relations warm up users. They predict based the preference of a user is closed to the average preference of its social relations. 
However, social relations are often incomplete or biased, leading to hard to predict accurate rating.

\subsection{HIN Construction for Cold-start Recommendation}
HIN-based recommender systems learn representations of users and items to model interactions. 
HeteMF \cite{hetemf} extracts meta-paths \cite{sun2011pathsim} to integrate item-item regularization with weighted MF and refine representation of users and items.
HeteRec \cite{heterec} leverages meta-path similarities to enrich user-item interaction matrix.
Hete-CF \cite{luo2014hete} utilizes heterogeneous relations and can be used in arbitrary social networks.
NIRec \cite{jin2020efficient} formulates interaction in a convolutional way and learn efficiently with fast Fourier transform.
NI-CTR \cite{neighbor} constructs four kinds of graphs and learns representations of neighbors. 
MetaHIN \cite{metahin} captures HIN-based semantics and leverage meta-learning to solve cold-start scenarios.
These works has to construct different heterogeneous interactions networks for different dataset, which is labor-intensive. Manually introduced interactions may lead to unreliable interactions.

Our method is totally different from these approaches. We do not need to inject prior knowledge to construct complex structure. Our framework is simple and generic, which can be applied for various datasets with different kinds of interactions.
}

%% file: preliminary.tex
\section{Preliminary}
\label{sec:preliminary}

We formulate the problem of cold-start rating prediction, and introduce multi-head self-attention based on which HIRE is developed.

\begin{figure}[t]
	\centering
  \resizebox{0.45\textwidth}{!}{
	\begin{tabular}[h]{c}
		\hspace{-0.3cm}
		\subfigure[User Cold-Start] {
			\includegraphics[width=0.3\columnwidth]{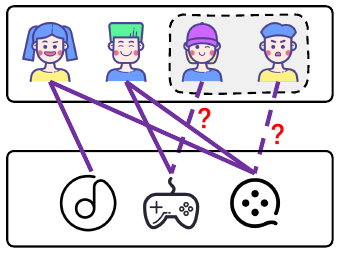}
			\label{fig:uc}
		}
		\subfigure[Item Cold-Start] {
			\includegraphics[width=0.3\columnwidth]{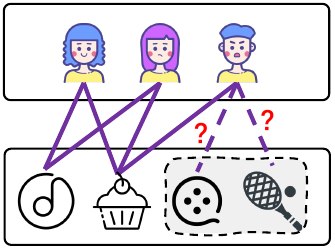}
			\label{fig:ic}
            }
             \subfigure[User\&Item Cold-Start] {
			\includegraphics[width=0.3\columnwidth]{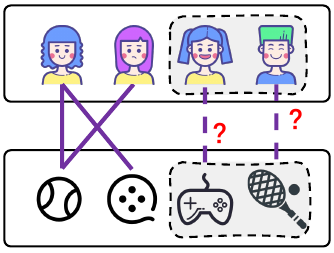}
			\label{fig:uic}
		} \\
	\end{tabular}}
	\caption{{3 cold-start scenarios for rating prediction: Entities in dotted boxes are the cold-start entities.}} 
        \label{fig:three}
\end{figure}

\subsection{Problem Statement}
\label{subsec:prob}
A recommendation system has a set of users $\UserSet = \{ u_1, \cdots, u_M \}$ and a set of items $\ItemSet = \{ i_1, \cdots, i_N \}$.
The users and items are associated with categorical attributes, i.e., the user profiles and item descriptions. 
Here, we use $\bm{e}^{k}_u$ and $\bm{e}^{k}_i$ to denote the $k$-th categorical attribute of a user $u$ and an item $i$, respectively, which are represented by one-hot encoding.
We also use $r_{ui}$ to denote the observed rating (preference) of user $u$ to item $i$.
Given a set of observed ratings, $\RateSet$, rating prediction is to predict unknown rating for pairs of users and items.

\begin{figure*}[t]
        \centering
	\includegraphics[width=0.9\textwidth]{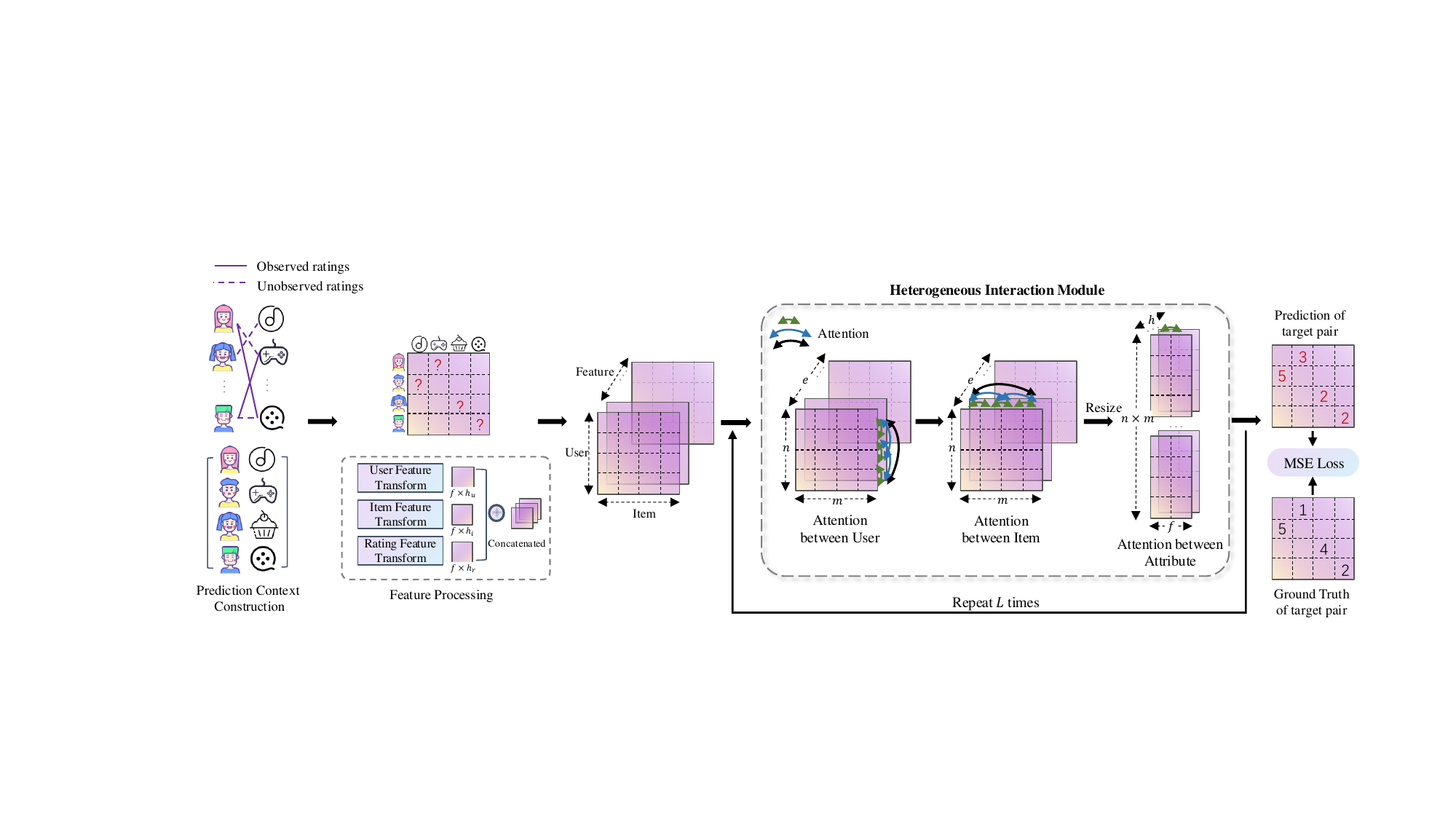} 
    \caption{\shadd{The architecture of HIRE:For ratings of cold user/items to be predicted, HIRE first samples a prediction context.  Then, the model constructs the initial embedding for the prediction context and learns the heterogeneous interactions via $L$ HIMs. Finally, the output embedding of the last HIM is transformed into the predicted rating matrix.}}
	\label{fig:CT}
\end{figure*}

\stitle{Cold-start rating prediction} is to predict cold-start users or/and items. 
There are 3 scenarios:
\begin{itemize}[leftmargin=*]
\item {{\sl User Cold-Start:} The user $u^*$ is new arrival, i.e., $u^* \notin \UserSet$, which has only a few rating interactions with existing items in $\ItemSet$.}
\item {{\sl Item Cold-Start:} The item $i^*$ is new arrival, i.e., $i^* \notin \ItemSet$, which has only a few rating interactions with existing users in $\UserSet$.}
\item {{\sl User\&Item Cold-Start:} Both the user $u^*$ and the item $i^*$ are new arrivals, i.e., $u^* \notin \UserSet$, $i^* \notin \ItemSet$.  They have only a few rating interactions with existing and other possible new items/users in the system.} 
\end{itemize}
Fig.~\ref{fig:three} shows an example of the 3 cold-start scenarios in a product rating system, where entities in the dotted boxes are cold-start entities. 
In Fig.~\ref{fig:uc}, it shows a User cold-start scenario to predict two new users' ratings on three existing items, in Fig.~\ref{fig:ic}, it shows an Item cold-start scenario to predict existing users' preference on two cold items (e.g., movie and tennis racket), and in Fig.~\ref{fig:uic}, it shows a User\&Item cold-start scenario to predict the ratings of two new users on 2 new items (e.g., game and tennis racket).

Given the user set $\UserSet$, item set $\ItemSet$ and the observed rating $\RateSet$, in this paper, we build a deep learning framework to support above cold-start rating prediction. It is important to note that by cold-start, the cold user and item to be predicted, as well as their associated ratings, are unavailable in the model training stage.

\subsection{Multi-Head Self-Attention}
\label{subsec:mhsa}

Multi-head self-attention (MHSA)~\cite{allyouneed} is the building block of foundation models, such as BERT~\cite{BERT}, GPT~\cite{brown2020language}, Vision Transformer~\cite{dosovitskiy2020image}, achieving high performance in various natural language processing and computer vision tasks.
Given a sequence of tokens $X = [x_1, \cdots, x_t]\in \mathbb{R}^{t \times d}$ as input, self-attention models the interactions among the tokens by computing the attention weights $A \in \mathbb{R}^{t \times t}$ as Eq.~\eqref{eq:att:linear}-\eqref{eq:att:weights}. \shadd{Here, $W_Q \in \mathbb{R}^{d \times d^k}$, $W_K \in \mathbb{R}^{d \times d^k}$ are linear weights that map $X$ to query matrix $Q \in \mathbb{R}^{t \times d^k}$ and key matrix $ K \in \mathbb{R}^{t \times d^k}$, respectively, $A_{ij}$ indicate the similarity between token $x_i$ and $x_j$, and $d^k$ and $d^v$ denote the dimensionality of the keys and values,} 
\begin{align}
Q = XW_Q, K = XW_K, \label{eq:att:linear} \\
A = \text{softmax}(\frac{QK^{T}}{\sqrt{d^k}}), \label{eq:att:weights}
\end{align}
Self-attention (SA) fuses the embedding of the input tokens by taking $A$ as the aggregation weights, where $W_V \in \mathbb{R}^{d \times d^v}$ is a linear weight matrix, 
\begin{align}
 \text{SA}(X) = A X W_V.   \label{eq:self-attention}
\end{align}
\shadd{MHSA computes $l_a$ self-attention independently and concatenates the results as Eq.~\eqref{eq:mhsa}, where the weight $W_O \in \mathbb{R}^{l_a\cdot d^v \times d^o}$ maps the concatenated dimension $l_a \cdot d^v$ to the output dimension $d^o$ and $\|$ is the concatenation operation,}
\begin{align}
    \text{MHSA}(X) = [\text{SA}^{(1)}(X) \| \cdots \| \text{SA}^{(l_a)}(X)] W_O. \label{eq:mhsa}
\end{align}
It is worth noting that MHSA is equivariant w.r.t. the permutation of the tokens.
Given a permutation function $\Pi_{[1:t]}$ that shuffles the $t$ tokens by an arbitrary order, we have:
\begin{align}
\Pi_{[1:t]} \circ \text{MHSA}(X) = \text{MHSA}(\Pi_{[1:t]} \circ X).
\end{align}

Our model adopts MHSA as the unique building block, which indicates the power of deep attention layers in modeling  heterogeneous relationships for recommendation tasks.

\comment{
\section{Preliminary}
\label{sec:preliminary}
\subsection{Problem Formulation}
\label{subsec:prob}
}
\comment{
\begin{table}[t]
\small
\centering
\caption{Notations}
\resizebox{0.4\textwidth}{!}{
\begin{tabular}{|c|c|}
\hline
Notation & Description                 \\\hline
 $U$/$I$  & user/item set in large dataset \\
$\mathcal{U}$/$\mathcal{I}$         & user/item set for a batch \\
 $n$/$d$ & number of users/items for a batch\\
 $Y$ & observed interactions existing in large dataset\\
 $\mathcal{R}$ & rating matrix for a batch \\
 $e_u$/$e_i$ & attributes of $u$/$i$\\
 $h_u$/$h_i$ & number of category attributes of $u$/$i$\\
 $\mathcal{E}$ & feature set for a batch\\
 $\mathcal{T}$ & a batch of data for input \\
 $\mathcal{S}$/$\mathcal{Q}$ & observed/masked ratings in a batch\\
  $\mathcal{D}$ & training batches\\
 $m_s$/$m_q$ & number of observed/masked ratings \\
 $f$ & embedding dimension for one category of feature \\
 $h$ & number of categories\\
 $e$ & embedding dimension for a pair of user and item\\
$f_U^s$ & user linear transformation for the $s$-th feature\\
$f_I^t$ & item linear transformation for the $t$-th feature\\
$f_R$ & rating linear transformation\\
 $\bm{H}$ & embedding matrix\\

 \hline
\end{tabular}}
\end{table}
}

\comment{
Given user set $U$, item set $I$ and their observed interactions $Y \in \mathbb{R}^{|U|\times|I|}$, where $y_{ui}=1$ denotes user $u$ have observed interactions with item $i$, otherwise $y_{ui}=0$. 
Users and items both include different category of features, e.g., users have $\{\text{age}, \text{gender}\}$ and items have $\{\text{publication year}, \text{director}\}$, etc. 
We use $e_u=\{e_{u}^1, e_{u}^2,\dots, e_u^{h_u}\}$ and $e_i=\{e_{i}^1, e_{i}^2,\dots, e_i^{h_i}\}$ to denote the features of user $u$ and item $i$,respectively. Let $h_u$ and $h_i$ denote the number of category features of user $u$ and item $i$. 

As input, we consider a batch of users $\mathcal{U}=\{u_1,u_2,\dots,u_n\}$ and a batch of items $\mathcal{I}=\{i_1,i_2,\dots,i_d\}$ which is a subset of $U$ and $I$, respectively. Let $|\mathcal{U}|=n$ and $|\mathcal{I}|=d$ denote the number of users and items in a subset, respectively. It has to be emphasized that $n$ and $d$ is flexible. User features and item features set are represented as $\mathcal{E}=\{e_u, e_i\}_{u\in \mathcal{U}, i \in \mathcal{I}}$.
The user preference for items are denoted as rating matrix $\mathcal{R} \in \mathbb{R}^{n\times d}$, where $r_{ui}$ denotes rating that user $u$ gives to item $i$. If there is no interaction between $u$ and $i$, we set $r_{ui}=-1$.
We define a batch of data as a input $\mathcal{T}=\{\mathcal{U}, \mathcal{I}, \mathcal{R}, \mathcal{E}\}$. 
The rating matrix $\mathcal{R}$ in a batch $\mathcal{T}$ contains a few observed ratings which we denote as $\mathcal{S}=\{r_{ui}\}_{m_s}$, and masked ratings as $\mathcal{Q}=\{r_{ui}\}_{m_q}$. $m_s$ is the number of ratings in $\mathcal{S}$, which can be a small value or even zero, and $m_q$ is the number of masked ratings in $\mathcal{Q}$, which aims to predict.

\stitle{Cold-start Recommendation.}
Our problem is to construct a model $\mathcal{M}$ to support cold-start rating prediction. The model is trained on a set of training batches $\mathcal{D}=\{\mathcal{T}_j\}_{j=1}^T$ and makes prediction on newly come test batches $\mathcal{T}^*=\{\mathcal{U}^*, \mathcal{I}^*, \mathcal{R}^*, \mathcal{E}^*\}$.
There are 3 possible cold-start scenarios from the perspective that user set and item set are constituted, including,
 \begin{itemize}[noitemsep,topsep=0pt,parsep=5pt,partopsep=0pt,leftmargin=*]
\item {\sl User-Cold-Start.} The users from test are new users of the model, i.e. $u \in \mathcal{U}^*$ does not appear in $\mathcal{U}$.
\item {\sl Item-Cold-Start.} The items from test are new items of the model, i.e., $i \in \mathcal{I}^*$ does not appear in $\mathcal{I}$.
\item {\sl User-and-Item-Cold-Start.} The users and items from test are both new for the model. $u \in \mathcal{U}^*$ does not appear in $\mathcal{U}$ and $i \in \mathcal{I}^*$ does not appear in $\mathcal{I}$.
\end{itemize}
To be specific, model $\mathcal{M}$ makes use of limited observed ratings in $\mathcal{S}$ of training batch to learn and model heterogeneous interactions. The parameters of model $\mathcal{M}$ are updated according to the prediction loss over $\mathcal{Q}$. 
In training phase, model $\mathcal{M}$ learns how to identify interactions between users, items and features and figure out significant parts to help rating predict.
 With a small size of ratings in $\mathcal{S}^*$ of test batch available, the model can swiftly make predictions by modeling heterogeneous interactions and learning from them. 

\subsection{Multi-Head Self-Attention}
\label{subsec:mhsa}

Multi-head self-attention (MHSA) has been widely applied in natural language process \cite{BERT, allyouneed}. Recently, research efforts extend the attention mechanism for tabular data \cite{transformer4tabular, annocolumn, NPT}. Multi-head self-attention allows the model to learn complex interactions between input sequence instances.
Attention function $\text{Att}(Q,K,V)$ maps queries $Q\in \mathbb{R}^{n\times h_k}$ to outputs using key-value pairs, the keys and values are packed together into matrices $K\in \mathbb{R}^{m\times h_k}$ and $V\in \mathbb{R}^{m\times h_v}$.
We compute dot-product attention,
\vspace{-1.5ex}
\begin{align}
\vspace{-1.5ex}
	\text{Att}(Q,K,V) &= \text{softmax}(QK^T/\sqrt{h})V. \label{eq:attention}
\end{align}
The pairwise dot product $QK^T$ measures the similarity between each pair of query and key vectors. The output is a weighted sum of $V$.
Instead perform a single dot product attention, a series of $k$ independent attention heads are concatenated, which allows the model to jointly attend to knowledge from different subspace. Multi-head attention is computed as follows,
\vspace{-1.5ex}
\begin{align}
\vspace{-1.5ex}
	\text{MHA}(Q,K,V) &= \text{concat}(O_1,\dots,O_k)W^O, \space \text{where} \label{eq:mhsa} \\
	O_j &= \text{Att}(QW_j^Q,KW_j^K,VW_j^V).
\end{align}
where $W_j^Q,W_j^K,W_j^V ,W^O$ are learnable parameters.
In this paper, we use multi-head self-attention, $\text{MHSA}(H)=\text{MHA}(Q=H,K=H,V=H)$, which set the same value for query $Q$, key $K$ and value $V$.
}

%% file: methodology.tex
\comment{
\begin{figure*}[t]
	\includegraphics[width=1\textwidth]{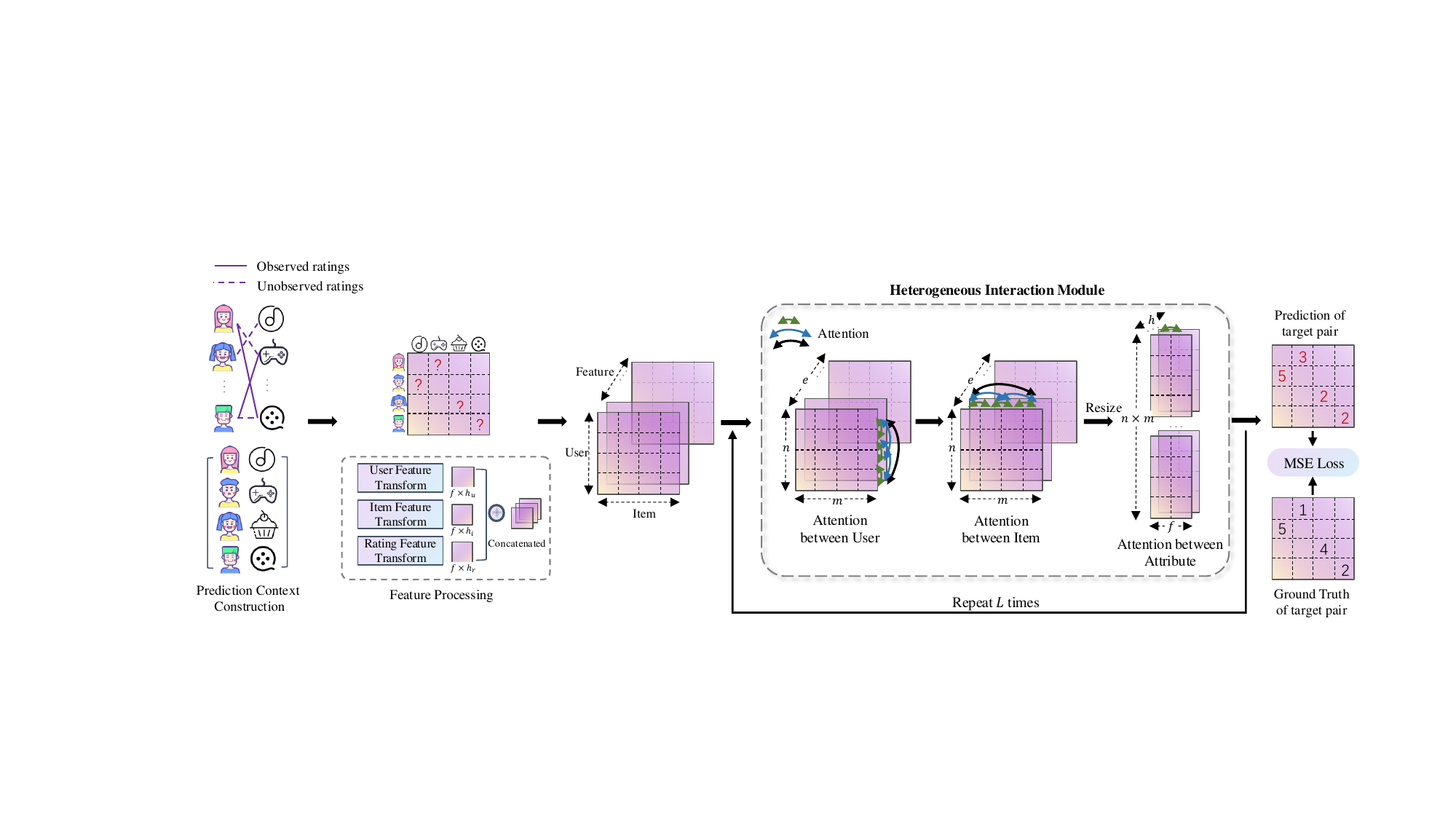} 
	\caption{\shadd{Architecture of HIRE. }Given a prediction context with unobserved ratings, we first perform linear transformation to get an embedding matrix. Then we apply $L$ \HIM blocks to model heterogeneous interactions. Finally we get a predicted matrix after mapping output of \HIM to ratings.}
	\label{fig:CT}
\end{figure*}}

\section{Methodology}
\label{sec:method}

We propose a DL framework, \underline{H}eterogeneous \underline{I}nteraction \underline{R}ating N\underline{e}twork (HIRE), for cold-start rating prediction. 
In this section, we present an overview of HIRE, followed by details on the input construction and the model architecture. 

\subsection{Overview}
\comment{
\shadd{
The basic idea of \HIRE is to design a fully data-driven model to learn heterogeneous interactions.
\HIRE models implicit or explicit interactions among a set of users or items, thus we develop a context prediction construction strategy to sample a set of users and items.
For different categories of user attributes, we input one-hot encodings and map them to dense embeddings via linear transformation. Take a female user for example, it has 'female' attribute and we denote the encoding as $[0,1]$. The one-hot encoding is mapped to a dense embedding. User embedding are concatenated by all categories of user dense embedding attributes. Item embedding are treated in a similar fashion. We transform all ratings to one-hot embeddings, and masked those unknown ratings, i.e., set them as all-zero vectors. Rating embedding is also mapped to a dense embedding via linear transformation.
When constructing embedding matrix, embedding for one user-item pair is generated by concatenating user embedding, item embedding and their corresponding rating embedding. 
It has to be noticed that all users and items in the set should be order independent. Ratings must be consistent even though the order of users or items are shuffled.}
}

\begin{figure}[t]
	\centering
	\includegraphics[width=0.9\columnwidth]{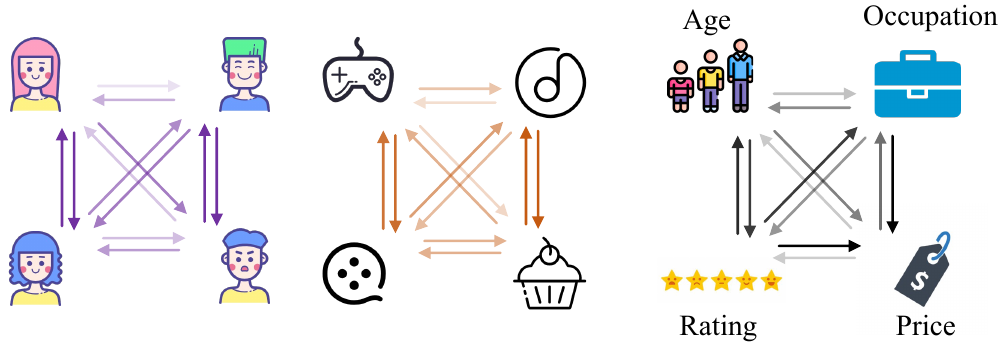}
	\vspace{-1ex}
	\caption{{Message Passing in complete graphs via learned interactions}}
        \label{fig:threeatten}
\end{figure}

The gist of \HIRE is to use a simple yet effective mechanism, MHSA, to holistically learn the interactions in three levels, i.e., users, items, and attributes, in an end-to-end, and data-driven fashion.
The overall architecture of \HIRE is illustrated in Fig.~\ref{fig:CT}. 
To predict the rating of a user on an item, where either the user or  item or both is cold-start, first we generate a prediction context which is composed of the target user/item, together with a set of pertinent users and items.   
These pertinent users and items are sampled from the user-item bipartite graph by leveraging the rating interaction. We envision that such users and items randomly selected by explicit rating relationships are informative for the cold-start user/item.  
Second, for a sampled prediction context, we construct a context matrix as the input of the model, which is a 3-dimensional tensor account for the users, items and attributes. The attributes contain individual attributes of users and items, and the observed ratings between the users and items in the context, where the ratings to be predicted are masked.  
\shadd{The context matrix is transformed by $L$ Heterogeneous Interaction Module (HIM) blocks, the main component of the \HIRE model, followed by a decoder to generate the prediction matrix.}
In each HIM block, there are three MHSA layers that learn the interactions between different users, different items and different categories of attributes, respectively.
Intuitively, as Fig.~\ref{fig:threeatten} shows, the three MHSA layers learn via message passing in three complete graphs, where the nodes are the users, items in the context, and the involved attributes, respectively. 
In the rest of this section, we will elaborate on the construction of the prediction context and the HIM in~\cref{subsec:context} and~\cref{subsec:modeldetail}, respectively.

From the perspective of the prediction task, what a \HIRE model does is analogous to inductive matrix completion~\cite{DBLP:conf/kdd/YingHCEHL18, DBLP:conf/iclr/ZhangC20}, where Graph Neural Networks are used to model the interactions in a matrix.
However, from the perspective of model design,  \HIRE is different from these GNN-based approaches. 
For one thing, GNN-based approaches only learn from a single type of rating relationship. For the other thing, the MHSA we use is a flexible specification of GNN layer~\cite{DBLP:journals/corr/abs-2202-08455}. 
As Eq.~\eqref{eq:att:weights}-\eqref{eq:self-attention} shows, self-attention conducts the message passing on a complete graph with a learned `soft' adjacency matrix instead of a fixed adjacency matrix.

\comment{
\shadd{
We give architecture of \HIRE in Fig.~\ref{fig:CT}. 
A set of users and items with a few of observed ratings are input. We construct an embedding matrix by three linear transformation. The main component is Heterogeneous Interaction Module (\HIM) and \HIM consists of three attention modules: Attention between users, attention between items and attention between attributes. There are $K$ \HIM blocks vertical stacked to learn high-order interactions between users, items, attributes, respectively. Finally, predictions are made via a MLP layer, mapping the output from \HIM to one-dimensional rating. We will elaborate the details of these components in \ref{subsec:modeldetail}.}
}

\subsection{Construction of Prediction Context}
\label{subsec:context}
In this section, we delve into the methodology of constructing the prediction context for cold users/items, which encompasses rich information for modeling the heterogeneous interactions to cold users/items.
For cold users/items, the prediction context comprises a set of users and items which are relevant to them, associated with their attributes and observed ratings. 
{
To determine the relevance of users and items, we employ a neighborhood-based sampling strategy to select the users and items from the user-item interaction bipartite graph. 
In a nutshell, given the limited budget of $n$ users and $m$ items in a prediction context, our sampling strategy preferentially samples those users/items that have rating interactions with the cold users/items.
Concretely, we initialize a seed set by the prediction target, i.e., pairs of users and items involving cold users/items.
Subsequently, the sampling begins at the seed set, by selecting the entities of the one-hop neighbors of the seed into the prediction context. If the number of the corresponding neighbor entities does not surpass the current budget, we put all the neighbor entities into the context. 
Otherwise, a subset of the neighborhood is sampled uniformly subject to the budget size.
}
The neighborhood sampling iterates hop by hop until the budget is exhausted. In~\cref{subsec:ablationstudy}, we also explore other sampling strategies to construct the context, such as random sampling and feature similarity sampling, where neighbourhood sampling achieves the best empirical results.
\begin{figure}[t]
\centering
\includegraphics[width=0.45\textwidth]{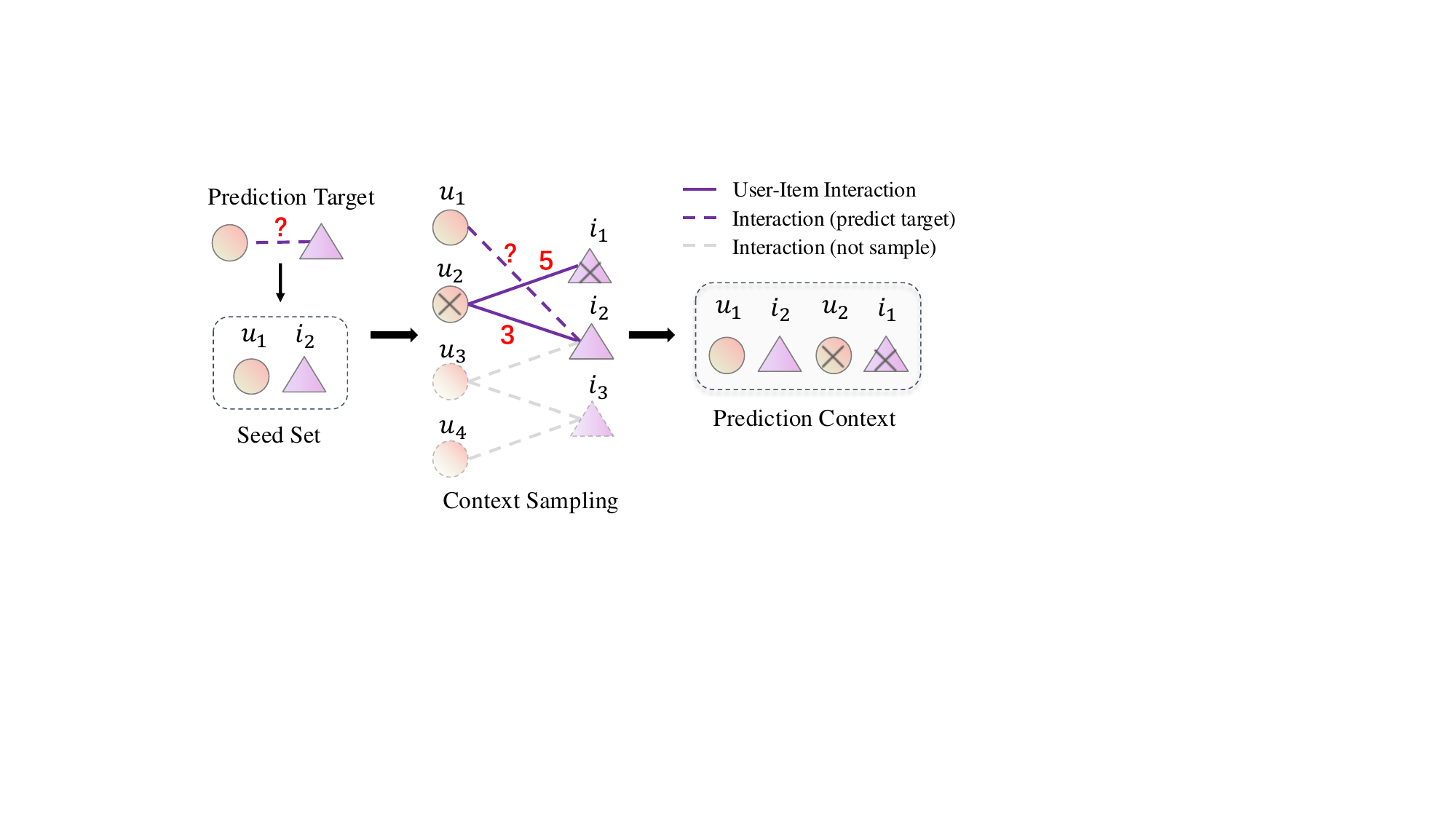} 
\vspace{-1ex}
\caption{\shadd{An example of prediction context construction}}	
	\label{fig:matrix}
    \vspace{-0.6cm}
\end{figure}
\comment{
Depending on the cold start scenario, the seed set contains specific types of nodes. For instance, in user-cold-start scenario, the seed set can be constructed by cold users with their interested items.
To construct a prediction context, we begin with selecting nodes from the seed set and appending them to corresponding user or item set. 
We randomly sample subsets of users and items from the one-hop neighborhood surrounding users and items in the seed set.
Then we consider to sample the remaining users (items) by maximizing their interactions with the sampled items (users).
}
Fig.~\ref{fig:matrix} shows a toy example of prediction context construction.

\begin{example}
\label{example:sample}
     Given $n = 2, m = 2$, we aim to predict the rating of a cold user $u_1$ on an item $i_2$. Initially, the seed set is fixed as $\{ u_1, i_2\}$, and an extra user and an extra item need to be sampled. In the bipartite rating graphs in Fig.~\ref{fig:matrix}, $i_2$ has two neighbors $\{u_2, u_3\}$, in which $u_2$ is sampled and the budget for users, $n$, is exhausted. As $u_1$ does not have any rated items, we can only sample an item from the neighbors of $u_2$, i.e., $i_1$. Finally, $\{ u_1, u_2, i_1, i_2\}$ forms the prediction context.   
\end{example}

After sampling $n$ users and $m$ items as the prediction context, we construct a 3-dimensional tensor as the initial input of the model. The tensor, denoted as $\bm{H} \in \Real^{n \times m \times e}$, encodes the potential attributes or identities of the users and items and the observed ratings. Concretely, in Eq.~\eqref{eq:concate}, the $k$-th row and $j$-th column of $\bm{H}$,  $\bm{H}[k, j, \cdot]$,  is an $e$-dimensional vector that concatenates three parts of the features, i.e., the features of the user $u_k$, $\bm{x}_{u_k}$, the features of the item $i_j$, $\bm{x}_{i_j}$, the vector representation of the potential rating of $u_k$ on $i_j$, $\bm{x}_r$. 
\begin{align}
 \bm{H}[k, j, \cdot] \leftarrow [\bm{x}_{u_k}  \| \bm{x}_{i_j} \| \bm{x}_r]. \label{eq:concate}
\end{align}
The user and item features derive from linear transformation from the original attributes into hidden space.
Let $\bm{e}_u^k$ denotes the one-hot embedding of the $k$-th categorical attribute of a user $u$. For the $h_u$ attributes of users, there are $h_u$ independent linear transformations $\{f_U^1, \cdots, f_U^{h_u}\}$ that transforms the one-hot embeddings into hidden features respectively, and the user feature $\bm{x}_u$ is the concatenation of those features as shown in Eq.~\eqref{eq:userfeats}. All the users in the recommendation system share the same set of linear feature transformations. 
And the item feature $\bm{x}_{i}$ is generated in a similar fashion as shown in Eq.~\eqref{eq:itemfeats}.
\begin{align}
 \bm{x}_u &\leftarrow [ f_U^1(\bm{e}_u^1) \| f_U^2(\bm{e}_u^2) \| \dots \| f_U^{h_u}(\bm{e}_u^{h_u}) ], \label{eq:userfeats} \\
 \bm{x}_i &\leftarrow [ f_I^1(\bm{e}_i^1) \| f_I^2(\bm{e}_i^2) \| \dots \| f_I^{h_i}(\bm{e}_i^{h_i}) ]. \label{eq:itemfeats}
\end{align}
For the ratings, we also use a linear transformation $f_R$ to map the concrete one-hot encoding of a rating $r$, $\bm{e}_r$, to the hidden representation $\bm{x}_r$ as shown in Eq.~\eqref{eq:rating},
\begin{align}
    \bm{x}_r \leftarrow f(\bm{e}_r),  \label{eq:rating}
\end{align}
if the rating $r$ is masked that will be predicted by the model, here, $\bm{e}_r$ and $\bm{x}_r$ are full-zero vectors. 
Suppose all the linear transformations $f_U^k$, $f_I^k$ and $f_R$ map all the one-hot embeddings into $f$-dimensional vectors. The dimension of the feature vector $\bm{H}[k, j, \cdot]$, $e$, equals to $(h_u + h_i + 1) \times f$.
If a user/item does not possess an attribute, we set $\bm{e}_u$/$\bm{e}_i$ to the one-hot encoding of the ID of the user/item as its unique attribute, which is also transformed by a linear mapping $f_U$/$f_I$.
The encoding of $\bm{H}$ can be easily extended to encode continuous attributes and ratings.

\comment{
\shadd{After generating user and item set, we construct $\mathcal{T}=\{\mathcal{U},\mathcal{I},\mathcal{R},\mathcal{E}\}$ with their attributes $\mathcal{E}$ and get $n\times d$ user-item pair. 
Their ratings $\mathcal{R}$ are divided into observed ratings $\mathcal{S}$ and masked ratings $\mathcal{Q}$.
Note if an user has no interaction with one item, we set a masked rating for this pair. }

\shadd{
User attributes contain different categories and each category are one-hot encoding.
User linear transformation $f_U^s: {e_u^s} \rightarrow x_u^s$ maps one-hot encoding to $f$-dimensional dense embedding, where $s$ refers to category of user attribute.
Given a user $u$ who has $h_u$ category of attributes, the embedding for the user is defined as:
\vspace{-1.5ex}
\begin{align}
\vspace{-1.5ex}
    x_u&=[f_U^1(e_u^1),f_U^2(e_u^2),\dots,f_U^{h_u}(e_u^{h_u})].
\end{align}
Similarly, item linear transformation $f_I^t: {e_v^t} \rightarrow x_i^t$ is utilized to map one-hot encoding to $f$-dimensional dense embedding. For an item $i$ with $h_i$ contents, the embedding for the item is defined as:
\vspace{-1.5ex}
\begin{align}
\vspace{-1.5ex}
    x_i&=[f_I^1(e_i^1),f_I^2(e_i^2),\dots,f_I^{h_i}(e_i^{h_i})].
\end{align}
It has to be noticed that the original embedding $e_u^s$ and $e_i^t$ are all one-hot encoding and they can have different dimensions, while $x_u^s$ and $x_i^t$ all have the same size, namely, $f$.
We input one-hot rating or masked rating (zero vector) into rating linear transformation $f_{R}: r_{ui} \rightarrow x_{ui}$,
\begin{align}
    x_{ui}=f_{R}(r_{ui}),
\end{align}
where $x_{ui}\in \mathbb{R}^f$.
Embedding $h_{ui}$ for a pair of user $u$ and item $i$ is concatenated by their user embedding, item embedding and rating embedding as shown,
\vspace{-1.5ex}
\begin{align}
\vspace{-1.5ex}
   {h}_{ui}&=[x_u,x_i,x_{ui}].\label{eq:concate}
\end{align}
Suppose $h=h_u+h_i+h_r$, where $h_u$ and $h_i$ denotes the number categories of attributes for user and item, respectively. And $h_r=1$ denotes that rating has one category, generally. Let $e$ denotes one pair embedding dimension, we can easily have $e=h \times f$.
Then for all the user-item pairs, we can get initial embedding matrix ${H} \in \mathbb{R}^{n \times d \times e}$ as input in heterogeneous interaction module.}

}

\subsection{Model Details}
\label{subsec:modeldetail}
We give the technical details of the heterogeneous interaction module (HIM) of HIRE. HIM fully exploits self-attention, a prevailing and powerful neural network layer, to learn implicit interactions. One HIM is composed of three different attention layers, i.e., attention between users, attention between items and attention between attributes, which are stacked layer by layer.

\stitle{{\underline{M}odeling interactions \underline{b}etween \underline{u}sers (\ABU).}} Given the initial embedding of the context $\bm{H} \in \Real^{n\times m \times e}$, where $n$ is the number of users, $m$ is the number of items, $e$ is the embedding dimension of attributes, we introduce an attention layer between users to model their implicit interactions. 
Specifically, we use $H_{i_j} = \bm{H}[\cdot, j, \cdot] \in \Real^{n\times e}$ to denote the embedding view of item $j$ in $\bm{H}$. 
An item embedding view, $H_{i_j}$, is processed by a multi-head self-attention (MHSA) layer, where the attention weights between $n$ users are computed, which is specific to item $j$. 
In Eq.~\eqref{eq:user:att1}, $\tilde{H}_{i_j} \in \Real^{m \times e}$ is the fused item embedding view weighted by the attention weights among users. 
All the item embedding views are  processed by a parameter-sharing MHSA in parallel, and the fused embeddings, $\{\tilde{H}_{i_j} | j \in [1, m] \}$, are stacked as the output embedding $\bm{H}^{(U)} \in \Real^{n \times m \times e}$ (Eq.~\eqref{eq:user:att2}),
\begin{align}
\tilde{H}_{i_j} &\leftarrow \text{MHSA}(H_{i_j}), \forall{j} \in [1, \cdots, m], \label{eq:user:att1}\\
\bm{H}^{(U)} &\leftarrow [ \tilde{H}_{i_1} \| \tilde{H}_{i_2} \|\cdots \| \tilde{H}_{i_m}]. \label{eq:user:att2}
\end{align}

\stitle{{\underline{M}odeling interaction \underline{b}etween \underline{i}tems (\ABI).}}
To further model the interaction between items, we introduce an attention layer between items, given the stacked output embedding $\bm{H}^{(U)} \in \Real^{n \times m \times e}$ as the input.
Similarly, ${H}_{u_k}=\bm{H}^{(U)}[k,\cdot,\cdot] \in \Real^{m\times e}$ denotes the embedding view of user $k$ in $\bm{H}^{(U)}$.
The attention weights between $m$ items are computed through a MHSA layer, which is specific to a user embedding view, $H_{u_k}$, indicating the user-specific interactions of items.
As Eq.~\eqref{eq:item:att1} shows, $\tilde{H}_{u_k}$ is the fused user embedding view weighted by the attention weights among items. A parameter-sharing MHSA processes $m$ independent user embedding views in parallel. The fused embeddings, $\{\tilde{H}_{u_k} | k \in [1, n] \}$, are stacked as the output embedding $\bm{H}^{(I)} \in \Real^{n \times m \times e}$ (Eq.~\eqref{eq:item:att2}),
\begin{align}
\tilde{H}_{u_k} &\leftarrow \text{MHSA}(H_{u_k}), \forall{k} \in [1, \cdots, n], \label{eq:item:att1} \\
\bm{H}^{(I)} &\leftarrow [ \tilde{H}_{u_1} \| \tilde{H}_{u_2} \|\cdots \| \tilde{H}_{u_n}].\label{eq:item:att2}
\end{align}

\stitle{{\underline{M}odeling interaction \underline{b}etween \underline{a}ttributes (\ABF).}}
HIM further learns the interaction between fine-grained attributes by computing the attention between user and item attributes.
Taking the embedding $\bm{H}^{(I)} \in \Real^{n \times m \times e}$ as input, we first reshape $\bm{H}^{(I)}$ to $\Real^{n \times m \times h \times f}$ by spliting the feature dimension $e$ into $h \times f$, where $h$ is the number of categorical attributes and $f$ is the embedding dimension for each attribute.
We use ${H}_{u_k, i_j}=\bm{H}^{(I)}[k,j,\cdot,\cdot] \in \Real^{h\times f}$ to denote the embedding view of a user-item pair $(u_k,i_j)$ in $\bm{H}^{(I)}$. The embedding view $H_{u_k,i_j}$ is processed by a MHSA layer, where the attention weights between $h$ attributes are computed, indicating the interaction between attributes that is specific to a user-item pair $(u_k,i_j)$.
$\tilde{H}_{u_k, i_j} \in \Real^{ h\times f}$ in Eq.~\eqref{eq:attr:att1} is the fused user-item pair embedding view weighted by the attention weights between the $h$ attributes. 
All the user-item pair embedding views are processed by a parameter-sharing MHSA in parallel, and the fused embeddings, $\{\tilde{H}_{u_k,i_j} | k \in[1,n],j \in [1, m] \}$, are stacked as the output embedding $\bm{H}^{(A)} \in \Real^{n \times m \times h\times f}$ (Eq.~\eqref{eq:attr:att2}).
The embedding $\bm{H}^{(A)}$ will be reshaped to $\Real^{n \times m \times e}$  by flatting the feature dimension $h \times f $ back to $e$ for future processing,
\begin{align}
\tilde{H}_{u_k, i_j} &\leftarrow \text{MHSA}(H_{u_k, i_j}), \forall{k} \in [1, \cdots, n],  \forall{j} \in [1, \cdots, m], \label{eq:attr:att1}
\end{align}
\begin{align}
\bm{H}^{(A)} \leftarrow [ \tilde{H}_{u_1, i_1} \| \tilde{H}_{u_1, i_2} \|\cdots \| \tilde{H}_{u_1, i_m} \nonumber \\
\tilde{H}_{u_2, i_1} \| \tilde{H}_{u_2, i_2} \|\cdots \| \tilde{H}_{u_2, i_m} \nonumber \\
\cdots \nonumber \\
\tilde{H}_{u_n, i_1} \| \tilde{H}_{u_n, i_2} \|\cdots \| \tilde{H}_{u_n, i_m} ]. \label{eq:attr:att2}
\end{align}
To summarize, HIM takes the sampled context $\bm{H} \in \Real^{n \times m \times e}$ as input embedding, and model the interactions between users, items and attributes by three MHSA layer by layer. It finally generates embedding $\bm{H}^{(A)} \in \Real^{n \times m \times e}$ which adaptively aggregates the features of users, items and attributes in the context by leveraging the attention weights.
\shadd{As Fig.~\ref{fig:CT} presents, the model HIRE is composed of $L$ HIMs, where the 
output of the $(l-1)$-th HIM is fed into $l$-th HIM for modeling the higher order interactions.}

\stitle{Rating Prediction.} Given $\bm{H}^{(A)} \in \Real^{n \times m \times e}$ as the output of the $L$-th HIM, we use a decoder to predict the rating matrix $\hat{R} \in \Real^{n \times m}$ of the sampled context as Eq.~\eqref{eq:predict},
\begin{align}
\hat{R} \leftarrow \alpha \cdot \text{Sigmoid}(g_{\theta}(\bm{H}^{(A)})). \label{eq:predict}
\end{align}
Here, $g_{\theta}: \Real^{e} \rightarrow \Real^{1}$ is a linear transformation parameterized by $\theta$ and $\alpha \in \Real$ is a scalar that rescales the range of the estimated ratings.

\section{Model Training and Analysis}
\label{sec:analysis}
In this section, we further present the training of \HIRE and analyze the complexity and inductive bias of the model. 
\subsection{Model Training}
In a recommendation system, given a set of users $\UserSet$,  a set of items $\ItemSet$ and observed ratings $\RateSet$, a \HIRE model is trained by a set of prediction contexts, denoted as $\mathcal{D}$, which are sampled from $\UserSet$, $\ItemSet$, and $\RateSet$. Given a ratio of $p$ ratings in a prediction context, the model is optimized to minimize the mean squared error (MSE) loss (Eq.~\eqref{eq:loss}) for the $1 - p$ of masked ratings,
\begin{align}
\loss= \frac{1}{|\mathcal{D}|} \sum_{\mathcal{D}} \frac{1}{|\mathcal{Q}|} \sum_{r  \in \mathcal{Q}} \norm{r-\hat{r}}^2, \label{eq:loss}
\end{align}
where $\hat{r} \in \Real$ is a predicted rating in the output matrix $\hat{R}$ and $r \in \Real$ is the corresponding ground truth rating. We use $\mathcal{Q}$ to denote the masked rating set. 

Algorithm~\ref{algo:train} presents the training process implemented by stochastic gradient descent. In each step, we randomly draw a mini-batch of prediction context, $\mathcal{B}$, from the training set $\mathcal{D}$ as line~\ref{line:sample}. 
Subsequently, in line~\ref{line:train:cube:start}-\ref{line:train:cube:end}, each context is transformed by the three attention layers in a HIM block one by one, and there are $L$ HIM blocks in total.
In line~\ref{line:train:decoder}, the output of the $L$-th HIM is mapped to the predicted rating matrix $\hat{R}$. 
Finally, we compute the MSE Loss for the mini-batch $\mathcal{B}$ and update the model parameters via the back-propagation of gradient in  line~\ref{line:train:loss}-\ref{line:train:update}.
The training will be terminated until the MSE loss converges. 

\comment{
\shadd{
In the training stage, given user set $\mathcal{U}$, item set $\mathcal{I}$ and observed ratings $\mathcal{R}$, \HIRE is trained by optimizing MSE Loss as follows:
\vspace{-1.5ex}
\begin{align}
\vspace{-1.5ex}
    \loss= \sum_{\mathcal{D}}\sum_{r_{ui} \in \mathcal{Q}} (r_{ui}-\hat{r}_{ui})^2, \label{loss}
\end{align}
where $\mathcal{Q}$ refers to masked rating set and $\mathcal{D}$ refers to a set of context matrix.
Algorithm 1 presents the training process.
First, we need to generate a set of context matrix $\mathcal{D}$ for training.
For each epoch (line~\ref{line:train:start}-\ref{line:train:end}), we sample a context matrix $\bm{H} \sim \mathcal{D}$. 
Second, let $\bm{H}$ go through $K$ \HIM blocks in line~\ref{line:train:cube:start}-\ref{line:train:cube:end}. In line~\ref{line:train:decoder}, we map $\bm{H}$ to one-dimensional rating $\hat{R}$. In line~\ref{line:train:loss}-\ref{line:train:update}, we compute MSE Loss for each $\hat{R}$, accumulate them and get loss $\mathcal{L}$ to update model parameters.
}
}

In the test stage, when new user set $\mathcal{U}^*$ and item set $\mathcal{I}^*$ arrives, we can construct an embedding matrix $\bm{H}^*$. Ratings to be predicted are masked as $0$.  
$\bm{H}^*$ is input into $L$ \HIM blocks following a mapping function to get predicted rating. 
\shadd{
\begin{example}
 Given the prediction context $\{ u_1, u_2, i_1, i_2\}$ in Example~\ref{example:sample}, we suppose both users and items have 2 attributes. 
 The context matrix $\bm{H}$ is a ${2\times 2 \times 80}$ tensor,  with each attribute and the rating are transformed into 16-dimensional vector by Eq.~\eqref{eq:userfeats}-\eqref{eq:rating}. 
 Specifically, the embedded ratings contain the observed rating of $(u_2, i_1)$ and $(u_2, i_2)$, the target rating of $(u_1, i_2)$ which is masked as $0$, and the unobserved rating of $(u_1, i_1)$ which is masked as $-1$. 
 Taking $\bm{H}$ as the model input, \ABU, \ABI, \ABF  compute the embeddings $\bm{H}^{(U)},\bm{H}^{(I)}, \bm{H}^{(A)}  \in \mathbb{R}^{2\times 2 \times 80}$, respectively.
 The process is repeated by $L$ times. 
 Finally, the last $\bm{H}^{(A)}$ is fed into a decoder in Eq.~\eqref{eq:predict} to generate the rating matrix $\hat{R} \in \mathbb{R}^{2 \times 2}$. The predicted rating of $(u_1, i_2)$ corresponds to the element in the first row and second column of the matrix, i.e., $\hat{R}[0,1]$.
\end{example}}

\subsection{Model Analysis}

We analyze the time and space complexity of \HIRE in brief. Suppose a prediction context is composed of $n$ users and $m$ items. $e = h \times f$ is the hidden dimension of the MHSA layer in HIM, where $h$ is the number of attributes and $f$ is the embedding dimension of each attribute. 
The linear feature transformation of \HIRE (Eq.~\eqref{eq:userfeats}-\eqref{eq:rating}) takes $\BigO(nmee_0)$ time,  where $e_0$ is the maximum dimension of original feature. 
The computation of HIM dominates the complexity of \HIRE. Specifically, the time complexities of the attention between users (Eq.~\eqref{eq:user:att1}-\eqref{eq:user:att2}), attention between items (Eq.~\eqref{eq:item:att1}-\eqref{eq:item:att2}) and attention between attributes (Eq.~\eqref{eq:attr:att1}-\eqref{eq:attr:att2}) are $\BigO(n^2me)$, $\BigO(nm^2e)$, $\BigO(nmh^2f)$, respectively. 
The time complexity of the output layer (Eq.~\eqref{eq:predict}) for rating prediction is $\mathcal{O}(nme)$. 
Assume $e_0 \ll e$ and the model is composed of $K$ HIM blocks, the total time complexity is $\BigO(Knme(n+m+h))$.
Regarding the space complexity, our neighborhood-based sampling strategy constructs a matrix which costs $\mathcal{O}(nme_0)$ space complexity for each task. For attention between users, items and attributes, the space complexity is $\mathcal{O}(n^2me)$, $\mathcal{O}(nm^2e)$ and $\mathcal{O}(nmh^2f)$, respectively. Thus the total space complexity for one context is $\mathcal{O}(Knme(n+m+h))$.

In addition, HIRE preserves an inherent inductive bias when exploiting sets of users and items for rating prediction, which is formulated as the property below. 

\begin{property}
\label{pro:sa}
Given a set of users $\{ u_1, \cdots, u_n\}$ and a set of items $\{ i_1, \cdots, i_m\}$ as a prediction context $\bm{H}$ and the HIRE model $\mathcal{M}$, the predicted rating matrix, $\hat{R}$, is  equivariant w.r.t. any permutation of the user set $\Pi_{[1:n]}$ and permutation of the item set $\Pi_{[1:m]}$,
\begin{align}
    \Pi_{[1:n]} \circ \Pi_{[1:m]} \circ \hat{R} = \mathcal{M}( \Pi_{[1:n]} \circ \Pi_{[1:m]} \circ \bm{H}).
\end{align}
\end{property}
The inductive bias is intuitive, since the MHSA is permutation equivariant., i.e., 
\begin{align}
   \Pi_{[1:n]} \circ \tilde{H}_{i_j} &\leftarrow \text{MHSA}( \Pi_{[1:n]} \circ H_{i_j}), \\
   \Pi_{[1:m]} \circ \tilde{H}_{u_k} &\leftarrow \text{MHSA}( \Pi_{[1:m]} \circ H_{u_k}),
\end{align}
and matrices stack/concatenation naturally preserve permutation equivariance.
{
This property ensures that given the sets of users and items in the prediction context,
our model predicts deterministic ratings regardless their
input order, which also shows that our approach can make use of MHSA designed for sequential inputs to solve the cold-start problem.
}

\comment{
\begin{algorithm}[t]
	\footnotesize
	\caption{Training Algorithm of \HIRE}
	\label{algo:train}
	\DontPrintSemicolon
	\SetKwData{Up}{up}  \SetKwInOut{Input}{Input} \SetKwInOut{Output}{Output}
    \SetKwRepeat{Do}{do}{while}
	\Input{User set $\mathcal{U}$, item set $\mathcal{I}$,  observed ratings $\mathcal{R}$.}
	\SetKwFunction{Emit}{Emit}
	\SetKwFunction{Check}{Check}

	Initialize model parameters; \\
    Generate a set of context matrix $\mathcal{D}$ for training; \\
        \Do{$\mathcal{L}$ converges}{ \label{line:train:start}
            Sample a context matrix $\bm{H} \sim \mathcal{D}$; \label{line:sample}\\
            \For {$k \rightarrow 1$ to $K$}{
                \label{line:train:cube:start}
                Compute $\bm{H}^{(U)}$ via Eq.~\eqref{eq:user:att1}-\eqref{eq:user:att2}; \label{line:train:cube:au}\\
                Compute $\bm{H}^{(I)}$ via Eq.~\eqref{eq:item:att1}-\eqref{eq:item:att2}; \label{line:train:cube:ai}\\
                Compute $\bm{H}^{(A)}$ via Eq.~\eqref{eq:attr:att1}-\eqref{eq:attr:att2}; \label{line:train:cube:af}\\
                \label{line:train:cube:end}
                }
            Compute the predicted rating matrix $\hat{R}$ via Eq.~\eqref{eq:predict};\label{line:train:decoder}\\
            Compute the MSE Loss $\mathcal{L}$ via Eq.~\eqref{loss}; \label{line:train:loss}\;
            Update model parameters by gradient descent; \label{line:train:update} \\
        } \label{line:train:end}
\end{algorithm}
}

\begin{algorithm}[t]
	\footnotesize
	\caption{Training Algorithm of \HIRE}
	\label{algo:train}
	\DontPrintSemicolon
	\SetKwData{Up}{up}  \SetKwInOut{Input}{Input} \SetKwInOut{Output}{Output}
    \SetKwRepeat{Do}{do}{while}
	\Input{User set $\mathcal{U}$, item set $\mathcal{I}$,  observed ratings $\mathcal{R}$.}
	\SetKwFunction{Emit}{Emit}
	\SetKwFunction{Check}{Check}

	Initialize model parameters; \\
    Generate a set of context matrices $\mathcal{D}=\{\bm{H}_1, \cdots \}$ for training; \\
        \Do{$\mathcal{L}$ converges}{ \label{line:train:start}
            Sample a mini-batch of context matrices $\mathcal{B} \sim \mathcal{D}$; \label{line:sample}\\
            \For {$\bm{H} \in \mathcal{B}$}{
            \For {$l \rightarrow 1$ to $L$}{
                \label{line:train:cube:start}
                Compute $\bm{H}^{(U)}$ via Eq.~\eqref{eq:user:att1}-\eqref{eq:user:att2}; \label{line:train:cube:au}\\
                Compute $\bm{H}^{(I)}$ via Eq.~\eqref{eq:item:att1}-\eqref{eq:item:att2}; \label{line:train:cube:ai}\\
                Compute $\bm{H}^{(A)}$ via Eq.~\eqref{eq:attr:att1}-\eqref{eq:attr:att2}; \label{line:train:cube:af}\\
                \label{line:train:cube:end}
                }
                Compute the predicted rating matrix $\hat{R}$ via Eq.~\eqref{eq:predict};\label{line:train:decoder}\\
            }
            Compute the MSE Loss $\mathcal{L}$ via Eq.~\eqref{eq:loss}; \label{line:train:loss}\;
            Update model parameters by gradient descent; \label{line:train:update} \\
        } \label{line:train:end}
\end{algorithm}

\comment{
\subsection{Heterogeneous Interaction Modeling}

Heterogeneous interaction module is composed of attention layers between users, items and features,respectively. We describe their details as follows.

\stitle{Modeling Interaction between Users.}
To model interaction between users, we introduce attention between user (\ABU).
 The \ABU layer computes attention between users for the same item which learns interactions between users explicitly. As input for \ABU, we reshape output of the previous layer as $d \times n\times e$. We take the number of items $d$ as batch size, applying MHSA independently to each column. The input embedding for each batch is denoted as $\mathcal{H}_i^{(l-1)}=\{h_{ui}\}_{u \in \mathcal{U}}\in \mathbb{R}^{n \times e}$,
 giving
\vspace{-1ex}
\begin{align}
\vspace{-1.5ex}
    \text{MHSA}(\mathcal{H}_i^{(l-1)})&=\mathcal{H}_i^{(l)} \in \mathbb{R}^{n\times e}, \label{ABU1}\\
    \text{\ABU}(\mathcal{H}^{(l)})&=\mathop{stack}\limits_{i\in \mathcal{I}} (\text{MHSA}(\mathcal{H}_{i_1}^{(l)}),MHSA(\mathcal{H}_{i_2}^{(l)}),\nonumber \\ 
    &\dots,\text{MHSA}(\mathcal{H}_{i_d}^{(l)})) \in \mathbb{R}^{d\times n\times e}.
\end{align}
The output from last layer
$\mathcal{H}^{(l-1)}$ is reshaped as ${d \times n \times e}$ and input in \ABU layer. The output of \ABU layer is $\mathcal{H}^{(l)} \in \mathbb{R}^{d \times n \times e}$.
Eq.~\ref{ABU1} computes MHSA amongst users for each batch, where attention mechanism models interactions amongst users. Intuitively, these operations are beneficial for estimating popularity of an item among users by weighting the importance of users.

\stitle{Modeling Interaction between Items.}
Attention between item layer (\ABI) are introduced to model interaction between items.
The \ABI layer computes attention between items for the same user which can explicitly reflects the essential parts that impact the user's preference.
As input for \ABI, the output of the previous layer is reshaped as $n \times d\times e$. Similar with \ABU, we take the number of users $n$ as batch size, and apply MHSA independently to each column. The embedding for each batch is denoted as $\mathcal{H}_u^{(l-1)}=\{h_{ui}\}_{i \in \mathcal{I}}\in \mathbb{R}^{d \times e}$, giving
\vspace{-1.5ex}
\begin{align}
\vspace{-1.5ex}
    \text{MHSA}(\mathcal{H}_u^{(l-1)})&=\mathcal{H}_u^{(l)} \in \mathbb{R}^{d\times e}, \label{ABI1}\\
    \text{\ABI}(\mathcal{H}^{(l)})&=\mathop{stack}\limits_{u\in \mathcal{U}} (\text{MHSA}(\mathcal{H}_{u_1}^{(l)}),MHSA(\mathcal{H}_{u_2}^{(l)}), \nonumber \\
    &\dots,\text{MHSA}(\mathcal{H}_{u_n}^{(l)})) \in \mathbb{R}^{n\times d\times e}.
\end{align}
The output of \ABI layer is $\mathcal{H}^{(l)} \in \mathbb{R}^{n\times d\times e}$. MHSA amongst items for a user is computed as Eq.~\ref{ABI1} shows to model interactions amongst items. It is helpful to estimate users' preference by weighting the importance of items.

\stitle{{Modeling Interaction between Features.}}
\ABU and \ABF model interactions from individual perspective, while neglect impact of fine-grained features. 
We propose attention between features (\ABF) to learn interactions between different categories features of one user-item pair. 
\ABF layer takes $n\times d$ user-item pairs as batch and split the embedding dimension $e$ into $h \times f$, where $h$ represents the number of categories for feature and $f$ represents embedding dimension for one category of feature. We apply MHSA independently to each pair, where the input of a pair denotes as $\mathcal{H}_{ui}^{(l-1)} \in \mathbb{R}^{h \times f}$, giving
\vspace{-1ex}
\begin{align}
\vspace{-1ex}
    \text{MHSA}(\mathcal{H}_{ui}^{(l-1)})&=\mathcal{H}_{ui}^{(l)} \in \mathbb{R}^{h\times f}, \label{ABF1}\\
    \text{\ABF}(\mathcal{H}^{(l)})&=\mathop{stack}\limits_{u\in \mathcal{U}, i\in \mathcal{I}} (\text{MHSA}(\mathcal{H}_{u_1i_1}^{(l)}),MHSA(\mathcal{H}_{u_1i_2}^{(l)}),\nonumber \\ 
    &\dots,\text{MHSA}(\mathcal{H}_{u_ni_d}^{(l)})) \in \mathbb{R}^{h\times f}.
\end{align}
The output of \ABF layer has the size of $[n\times d, h, f]$. Eq.~\ref{ABF1} computes MHSA amonst all categoreis of features for a certain pair. This module can reflect impact of each category of feature on rating.
By stacking above three layers, a heterogeneous interaction module is composed. After vertical stacking multiple \HIM blocks, \CT can learn higher-order different pattern of interactions between users, items and features.

To obtain a predict rating matrix, we use a MLP layer mapping output of last \HIM block $\mathcal{H} \in \mathbb{R}^{n \times d \times e}$ to $\hat{R}\in \mathbb{R}^{n\times d}$, as Eq.~\ref{eq:decoder} shows,
\vspace{-1ex}
\begin{align}
\vspace{-1ex}
    \hat{R} = \alpha\sigma(g_{\theta}(\mathcal{H})), \label{eq:decoder}
\end{align}
where $\sigma$ is a Sigmoid function, $\alpha$ is a scaler to convert the probability to ratings and $g_\theta$ is a MLP layer which maps dimension $e$ to $1$.

 Importantly, \HIM are equivariant to a permutation of users or items. In other words, if set of users or items is shuffled, \HIM can still produce the same rating prediction shuffled in a same manner. It requires interaction modeling should not depend on ordering of these users or items. We give a brief proof of permutation equivariant for \HIM in Section~\ref{sec:pro}.
}

\comment{
\begin{algorithm}[t]
	\footnotesize
	\caption{Training Algorithm of \CT}
	\label{alg}
	\DontPrintSemicolon
	\SetKwData{Up}{up}  \SetKwInOut{Input}{Input} \SetKwInOut{Output}{Output}
	\Input{User set $U$, item set $I$, features, interaction matrix $Y$.}
	\Output{All model parameters.}
	\SetKwFunction{Emit}{Emit}
	\SetKwFunction{Check}{Check}

	Initialize all model parameters.\;
        Generate training batches $D\{\mathcal{T}_j\}_{j=1}^T$ where each batch contains $\mathcal{T}=\{\mathcal{U}, \mathcal{I}, \mathcal{R}, \mathcal{E}\}$ using strategy in Section~\ref{generate};\;
		\For {$epoch \rightarrow 1$ to $L$}
		{ \label{line:train:start}
                
			\For { $i\rightarrow 1$ to $T$ }
			{ \label{line:train:task:start}
    Construct input matrix $\mathcal{H}(i) \in \mathbb{R}^{n \times d\times e}$ for a batch by generating embeddings $h_{ui}$ via linear transformation $f_U$, $f_I$ and $f_R$ using \cref{eq:concate}. \label{line:train:encoder}\label{encoder}\;
    \For {$k \rightarrow 1$ to $K$}
    {
    \label{line:train:cube:start}
        Resize $\mathcal{H}(i)$ as $[d,n,e]$ size.\;
        Calculate $\mathcal{H}(i) \leftarrow$ \ABU$(\mathcal{H}(i))$.\label{line:train:cube:au}\;
        Resize $\mathcal{H}(i)$ as $[n,d,e]$ size.\;
        Calculate $\mathcal{H}(i) \leftarrow$ \ABI$(\mathcal{H}(i))$.\label{line:train:cube:ai}\;
        Resize $\mathcal{H}(i)$ as $[n\times d,h,f]$ size.\;
        Calculate $\mathcal{H}(i) \leftarrow$ \ABF$(\mathcal{H}(i))$.\label{line:train:cube:af}\;
    \label{line:train:cube:end}
    }
    Map $\mathcal{H}(i)$ to one-dimensional rating $\hat{R}(i)$ using \cref{eq:decoder}.\label{line:train:decoder}\;
    
\label{line:train:task:end}			}
   Calculate MSELoss $\mathcal{L}$ by \cref{loss}. \label{line:train:loss}\;
   Update model parameters by optimizer. \label{line:train:update}\
   \label{line:train:end}}
\Return{All model parameters}.\
\end{algorithm}
}

\comment{
\subsection{Proof of Permutation Equivariant}
\label{sec:pro}
Here we give a brief proof that \HIM is permutation equivariant.
\begin{property}
\label{pro:sa}
  Heterogeneous Interaction Module (\HIM) is equivariant to permutation of users and items.   
\end{property}

\myproof
    Let $X\in \mathbb{R}^{d \times e}$ and $W \in \mathbb{R}^{e\times e_2}$, we first prove $XW$ is row-equivariant. Suppose $\pi(X)$ be a permutation of the rows of $X$, then 
    \vspace{-1.5ex}
    \begin{align}
    \vspace{-1.5ex}
    (\pi(X) W)_{ij} &= \sum_k \pi X_{ik}W_{kj}\label{eq:rowequivariant} \\
    &=\sum_k X_{\pi(i)k}W_{kj} \nonumber \\
    &=(XW)_{\pi(i)j}\nonumber \\
    &=\pi(XW)_{ij}. \nonumber  
    \end{align}
    Since we have
    \vspace{-1.5ex}
    \begin{align}
    \vspace{-1.5ex}
    \text{Att}(XW^Q,XW^K,XW^V)=\sigma(XW^Q(XW^K)^\top/\sqrt{h})XW^V,
    \end{align}
    where $\sigma$ is denoted as row-wise softmax function. Then
    \vspace{-1.5ex}
    \begin{align}
    \vspace{-1.5ex}
(\pi X W^Q(\pi X W^K)^\top)_{ij} &=\pi(XW^Q)\pi(XW^K)^\top_{jk} \label{eq:attequivariant}\\
&=\sum_k (\pi(XW^Q))_{ik}(\pi(XW^K))_{jk} \nonumber \\
&=\sum_k XW^Q(XW^K)^\top_{\pi(i)\pi(j)}\nonumber.
    \end{align}
    Let $M=\sigma(XW^Q(XW^K)^\top/\sqrt{h}$, 
    \vspace{-1ex}
    \begin{align}
    \vspace{-1ex}
(\pi \sigma(XW^Q(XW^K)^\top/\sqrt{h})(\pi XW^V))_{ij} &=\pi (M)(\pi XW^V)_{ij}\label{eq:saequivariant}\\
&=\pi(M) \pi(XW^V)_{ij} \nonumber\\
&=\sum M_{\pi(i)\pi(k)}(XW^V)_{\pi(k)j} \nonumber\\
&=M(XW^V)_{\pi(i)j}, \nonumber
    \end{align}
now we show self-attention is row-equivariant \cite{NPT}.
We then prove \ABI layer is permutation equivariant for users and items. 
For the input $H\in \mathbb{R}^{n\times d\times e}$, $n$ is a batch size, multi-head self-attention is applied to each row independently, which is not related to the order of users. Then we have permutation equivariant for users.
We consider $h_u \in \mathbb{R}^{d \times e}$ for each row, and easily get
\vspace{-1ex}
    \begin{align}
    \vspace{-1ex}
        \text{MHSA}(\pi H_u)=\pi \text{MHSA}(H_u),
    \end{align}
    indicating permutation equivariant of items is established. We get permutation equivariant of users and items in \ABI layer.
    Similarly, we can easily get \ABU layer has same property as \ABI layer.
    For \ABF layer, since input is reshaped as $H \in \mathbb{R}^{n\times d \times h\times f}$ where $n \times d$ is batch size. We can immediately get the order of users and items is irrelevant with the model. Thus we have \ABF layer satisfying permutation equivariant for both users and items.
    \HIM are vertical stacked by \ABU, \ABI and \ABF layers, we can finally come to a conclusion that \HIM is equivariant to permutation of users and items.
\eop
}

\comment{
\subsection{Complexity Analysis}
\shadd{We analyze the time complexity of \CT in brief. 
For linear transformation of \CT, the time complexity is $\mathcal{O}(nmee_0)$ for one task, where $e_0$ is the maximum original feature dimension. 
The main time cost is heterogeneous interaction module. For \ABU layers, the computation complexity is $\mathcal{O}(n^2me)$. For \ABI layers, the computation complexity is $\mathcal{O}(nm^2e)$ and the computation complexity for \ABF layers is $\mathcal{O}(nmh^2f)$. 
The time complexity of MLP layer for rating prediction is $\mathcal{O}(nme)$. 
Assume $e_0$ is less than $e$, the total cost for one task is $\mathcal{O}(Knme(n+m+h))$, where $K$ is the number of stacked layers.
Regarding the space complexity, our 2-hop sampling strategy construct a matrix which costs $\mathcal{O}(nme_0)$ space complexity for each task. For \ABU, \ABI and \ABF layers, the space complexity is $\mathcal{O}(n^2me)$, $\mathcal{O}(nm^2e)$ and $\mathcal{O}(nmh^2f)$, respectively. Thus the total space complexity for one task is $\mathcal{O}(Knme(n+m+h))$.
}
}

%% file: experiment.tex
\begin{table}[t]
\caption{Profile of Datasets}
\label{tab:dataset}
\centering
\footnotesize
\resizebox{0.5\textwidth}{!}{
\begin{tabular}{c|r|r|r}
\toprule
Dataset           & \Movielens                                                             & \Douban         &\Bookcrossing  \\\midrule
\# Users   & 6,040                                                                        &   23,822           & 278,858   \\
\# Items   & 3,706                                                                        &  185,574            & 271,379  \\
\# Ratings & 1,000,209                                                                      &  1,387,216             &  1,149,780  \\\midrule
User Attributes     & \makecell[r]{Age, Occupation, \\ Gender, Zip code} &  N/A                 & Age  \\\midrule
Item Attributes     & \makecell[r]{Rate, Genre, \\Director, Actor} & N/A      &Publication year  \\\midrule
Range of Ratings  & $1 \sim 5$ & $1 \sim 5$ & $1 \sim 10$ \\
\bottomrule
\end{tabular}
}
\end{table}

\begin{table*}[t]
\footnotesize 
\centering
\caption{{Overall Performance in Warm States (\%)}}
\label{tab:result:warmstate}
\resizebox{1.0\textwidth}{!}{
\begin{tabular}{c|c|ccc|ccc|ccc}
\toprule
\multirow{2}{*}{Datasets}                & \multirow{2}{*}{Methods} & \multicolumn{3}{c|}{TOP5}                                           & \multicolumn{3}{c|}{TOP7}                                           & \multicolumn{3}{c}{TOP10}                                          \\ 
 &                          & Precision            & NDCG                 & MAP                  & Precision            & NDCG                 & MAP                  & Precision            & NDCG                 & MAP                  \\ \midrule 
\multirow{4}{*}{\Movielens}  
& \Popularity       & $57.63_{\small(0.52)}$ & $86.02_{\small(0.13)}$ & $47.41_{\small(0.21)}$ & $62.47_{\small(0.63)}$ & $86.64_{\small(0.11)}$ & $51.74_{\small(0.38)}$ & $69.69_{\small(0.63)}$ & $87.80_{\small(0.08)}$ & $59.63_{\small(0.58)}$\\
& \NEUMF       & $49.92_{\small(0.12)}$ & $73.01_{\small(0.11)}$ & $40.83_{\small(0.10)}$ & $52.25_{\small(0.19)}$ & $74.97_{\small(0.15)}$ & $42.26_{\small(0.16)}$ & $56.01_{\small(0.20)}$ & $76.62_{\small(0.17)}$ & $45.71_{\small(0.21)}$\\
& \TANP       & $58.73_{\small(0.12)}$ & $87.24_{\small(0.06)}$ & $48.66_{\small(0.04)}$ & $61.79_{\small(0.09)}$ & $87.55_{\small(0.05)}$ & $50.95_{\small(0.20)}$ & $67.21_{\small(0.17)}$ & $88.27_{\small(0.02)}$ & $56.56_{\small(0.24)}$\\
& \HIRE       & 
\cellcolor{LightYellow}$\mathbf{78.23_{\small(0.57)}}$ & \cellcolor{LightYellow}$\mathbf{95.86_{\small(0.26)}}$ & \cellcolor{LightYellow}$\mathbf{73.06_{\small(0.68)}}$ & \cellcolor{LightYellow}$\mathbf{78.46_{\small(0.37)}}$ & \cellcolor{LightYellow}$\mathbf{96.58_{\small(0.24)}}$ & \cellcolor{LightYellow}$\mathbf{73.60_{\small(0.46)}}$ & \cellcolor{LightYellow}$\mathbf{78.51_{\small(0.13)}}$ & \cellcolor{LightYellow}$\mathbf{98.03_{\small(0.22)}}$ & \cellcolor{LightYellow}$\mathbf{73.90_{\small(0.16)}}$
\\
\hline           
\end{tabular}
}
\end{table*}

\begin{table*}[t]
\footnotesize 
\centering
\caption{\shadd{Overall Performance in 3 Cold-Start Scenarios on \Movielens (\%)}}
\label{tab:result:movielens}
\resizebox{1.0\textwidth}{!}{
\begin{tabular}{c|c|ccc|ccc|ccc}
\toprule 
\multirow{2}{*}{Scenarios}                & \multirow{2}{*}{Methods} & \multicolumn{3}{c|}{TOP5}                                           & \multicolumn{3}{c|}{TOP7}                                           & \multicolumn{3}{c}{TOP10}                                          \\ 
 &                          & Precision            & NDCG                 & MAP                  & Precision            & NDCG                 & MAP                  & Precision            & NDCG                 & MAP                  \\ \midrule 
\multirow{13}{*}{\UC}        
 & \shadd{\Popularity}                  &\shadd{\underline{$58.20_{\small (0.11)}$} }& 
\shadd{$85.40_{\small (0.06)}$}  & \shadd{\underline{$47.38_{\small (0.16)}$}} & \shadd{\underline{$64.07_{\small (0.13)}$}} & \shadd{$86.40_{\small (0.05)}$} & \shadd{\underline{$52.95_{\small (0.17)}$}} & \shadd{\underline{$72.89_{\small (0.15)}$}} &\shadd{{$88.00_{\small (0.06)}$}}  & \shadd{\underline{$63.38_{\small (0.20)}$}}                \\
 & \shadd{\transgnn}                & 
\shadd{$55.20_{\small (1.80)}$} & \shadd{$56.86_{\small (1.69)}$}  & \shadd{$41.29_{\small (2.11)}$} & \shadd{$56.39_{\small (1.80)}$} & \shadd{$57.33_{\small (2.06)}$} & \shadd{$41.40_{\small (2.79)}$} & \shadd{$58.12_{\small (0.48)}$} &\shadd{$57.50_{\small (1.57)}$}  & \shadd{$43.40_{\small (1.48)}$}                \\
\bottomrule 
\end{tabular}}
\end{table*}

\begin{table*}
\footnotesize 
\centering
\caption{\shadd{Overall Performance in 3 Cold-Start Scenarios on \Bookcrossing (\%)}}
\label{tab:result:bookcross}
\resizebox{1.0\textwidth}{!}{
\begin{tabular}{c|c|ccc|ccc|ccc}
\toprule 
\multirow{2}{*}{Scenarios}                & \multirow{2}{*}{Methods} & \multicolumn{3}{c}{TOP5}                                           & \multicolumn{3}{|c}{TOP7}                                           & \multicolumn{3}{|c}{TOP10}                                          \\ 
 &                          & Precision            & NDCG                 & MAP                  & Precision            & NDCG                 & MAP                  & Precision            & NDCG                 & MAP                  \\\midrule 
\multirow{11}{*}{\UC} 
 & \shadd{\Popularity}                  &\shadd{{$37.68_{\small (1.07)}$} }& 
\shadd{$83.77_{\small (0.39)}$}  & \shadd{$26.33_{\small (1.55)}$} & \shadd{$42.33_{\small (1.74)}$} & \shadd{$84.14_{\small (0.19)}$} & \shadd{{$27.51_{\small (1.69)}$}} & \shadd{$48.01_{\small (0.14)}$} &\shadd{$86.10_{\small (0.21)}$}  & \shadd{$31.29_{\small (0.95)}$}                \\
 & \shadd{\transgnn}                & 
\shadd{$38.75_{\small (1.76)}$} & \shadd{$41.43_{\small (1.72)}$}  & \shadd{$32.11_{\small (1.56)}$} & \shadd{$41.07_{\small (2.52)}$} & \shadd{$43.44_{\small (2.27)}$} & \shadd{$34.34_{\small (2.35)}$} & \shadd{$45.01_{\small (0.07)}$} &\shadd{$46.24_{\small (0.67)}$}  & \shadd{$36.83_{\small (1.53)}$}                \\
& NeuMF      & $33.28_{\small(1.06)}$ & $38.87_{\small(1.10)}$  & $26.57_{\small(1.67)}$ & $41.76_{\small(1.12)}$ & $39.62_{\small(1.32)}$ & $33.88_{\small(2.03)}$ & $51.65_{\small(1.55)}$ & $41.78_{\small(1.27)}$ & $42.14_{\small(2.71)}$ \\
\hline        
\end{tabular}
}
\end{table*}

\comment{
\begin{table*}
\footnotesize 
\centering
\caption{{Overall Performance in 3 Cold-Start Scenarios on \Movielens (\%)}}
\label{tab:result:movielens}
\resizebox{0.9\textwidth}{!}{
\begin{tabular}{c|c|ccc|ccc|ccc}
\toprule 
\multirow{2}{*}{Scenarios}                & \multirow{2}{*}{Methods} & \multicolumn{3}{c|}{TOP5}                                           & \multicolumn{3}{c|}{TOP7}                                           & \multicolumn{3}{c}{TOP10}                                          \\ 
 &                          & Precision            & NDCG                 & MAP                  & Precision            & NDCG                 & MAP                  & Precision            & NDCG                 & MAP                  \\ \midrule 
\multirow{10}{*}{\UC}          & NeuMF                  & 
$47.02_{\small (0.66)}$ &$67.13_{\small (0.52)}$  & $37.13_{\small (0.68)}$ & $50.26_{\small (0.67)}$ & $68.99_{\small (0.45)}$ & $39.47_{\small (0.63)}$ & $54.78_{\small (0.65)}$ &$70.73_{\small (0.34)}$  & $43.91_{\small (0.68)}$                \\
& Wide\&Deep               &
$51.89_{\small (0.83)}$ & $83.85_{\small (0.15)}$ & $41.57_{\small (1.02)}$ & $55.61_{\small (0.47)}$ & $83.96_{\small (0.21)}$ & $44.20_{\small (0.58)}$ & $62.48_{\small (0.66)}$ & $84.71_{\small (0.15)}$ & $50.81_{\small (0.93)}$ \\
& DeepFM &
$51.69_{\small (0.92)}$ & $83.67_{\small (0.79)}$ & $41.23_{\small (0.93)}$ & $55.36_{\small (0.53)}$ & $83.85_{\small (0.73)}$ & $43.84_{\small (0.62)}$ & $61.90_{\small (0.25)}$ & $84.51_{\small (0.63)}$ & $50.19_{\small (0.28)}$ \\
& AFN &$50.84_{\small (0.57)}$ & $82.94_{\small (0.32)}$ & $39.98_{\small (0.56)}$ & $55.50_{\small (0.33)}$  & $83.33_{\small (0.29)}$ & $43.40_{\small (0.33)}$  & $62.54_{\small (0.26)}$ & $84.23_{\small (0.18)}$ & $50.28_{\small (0.43)}$\\
&GraphHINGE &$51.80_{\small (0.75)}$  & $77.75_{\small (0.69)}$ & $40.76_{\small (0.71)}$ & $58.19_{\small (0.86)}$ & $78.06_{\small (0.82)}$ & $46.55_{\small (0.86)}$ & $64.62_{\small (0.76)}$ & $78.26_{\small (0.79)}$ & $53.03_{\small (0.94)}$\\
\bottomrule 
\end{tabular}}
\end{table*}}

\comment{
\begin{table*}
\footnotesize 
\centering
\caption{Overall Performance in Three Cold-Start Scenarios on \Bookcrossing (\%)}
\label{tab:result:bookcross}
\resizebox{0.9\textwidth}{!}{
\begin{tabular}{c|c|ccc|ccc|ccc}
\toprule 
\multirow{2}{*}{Scenarios}                & \multirow{2}{*}{Methods} & \multicolumn{3}{c}{TOP5}                                           & \multicolumn{3}{|c}{TOP7}                                           & \multicolumn{3}{|c}{TOP10}                                          \\ 
 &                          & Precision            & NDCG                 & MAP                  & Precision            & NDCG                 & MAP                  & Precision            & NDCG                 & MAP                  \\\midrule 
\multirow{7}{*}{\UC}          & NeuMF      & $33.28_{\small(1.06)}$ & $38.87_{\small(1.10)}$  & $26.57_{\small(1.67)}$ & $41.76_{\small(1.12)}$ & $39.62_{\small(1.32)}$ & $33.88_{\small(2.03)}$ & $51.65_{\small(1.55)}$ & $41.78_{\small(1.27)}$ & $42.14_{\small(2.71)}$ \\
 & Wide\&Deep & $28.52_{\small(1.61)}$ & $54.08_{\small(1.97)}$ & $21.61_{\small(0.08)}$ & $37.59_{\small(2.31)}$ & $55.28_{\small(2.73)}$ & $25.17_{\small(1.96)}$ & $47.46_{\small(6.78)}$ & $56.66_{\small(4.16)}$ & $32.58_{\small(5.04)}$ \\
 & DeepFM     & $29.56_{\small(0.13)}$ & $51.16_{\small(0.65)}$ & $18.70_{\small(0.79)}$ & $35.11_{\small(1.25)}$ & $51.43_{\small(0.57)}$ & $21.20_{\small(0.70)}$ & $43.21_{\small(2.78)}$ & $52.15_{\small(0.18)}$ & $27.87_{\small(2.33)}$ \\
 & AFN        & $22.05_{\small(7.14)}$ & $49.70_{\small(3.54)}$ & $14.62_{\small(6.44)}$ & $31.95_{\small(4.14)}$ & $50.63_{\small(2.64)}$ & $18.56_{\small(4.52)}$ & $42.10_{\small(0.66)}$ & $52.38_{\small(0.80)}$ & $25.73_{\small(2.77)}$ \\
 & MAMO     &$40.16_{\small(1.15)}$ & $27.52_{\small(1.94)}$ & $30.62_{\small(1.23)}$ & \underline{$54.44_{\small(0.70)}$} & $32.44_{\small(1.84)}$ & \underline{$44.19_{\small(0.91)}$} & \underline{$71.19_{\small(0.46)}$} & $38.95_{\small(1.50)}$  & \underline{$64.17_{\small(0.65)}$} \\
\hline        
\end{tabular}
}
\end{table*}}

\comment{
\begin{table*}
\footnotesize 
\centering
\caption{{Overall Performance in Three Cold-Start Scenarios on \Douban (\%)}}
\label{tab:result:douban}
\resizebox{0.9\textwidth}{!}{
\begin{tabular}{c|c|ccc|ccc|ccc}
\toprule
\multirow{2}{*}{Scenarios}                & \multirow{2}{*}{Methods} & \multicolumn{3}{c|}{TOP5}                                           & \multicolumn{3}{c|}{TOP7}                                           & \multicolumn{3}{c}{TOP10}                                          \\ 
 &                          & Precision            & NDCG                 & MAP                  & Precision            & NDCG                 & MAP                  & Precision            & NDCG                 & MAP                  \\ \midrule 
\multirow{8}{*}{\UC}      
& NeuMF      & $44.43_{\small(1.60)}$  & $33.34_{\small(1.99)}$ & $40.56_{\small(1.64)}$ & $58.72_{\small(1.24)}$ & $34.71_{\small(1.91)}$ & $56.23_{\small(1.30)}$ & $72.61_{\small(0.71)}$ & $37.65_{\small(1.68)}$ & $71.13_{\small(0.80)}$    \\
 & Wide\&Deep & $54.42_{\small(1.05)}$ & $73.76_{\small(0.70)}$ & $44.43_{\small(1.75)}$ & $55.66_{\small(0.31)}$ & $75.95_{\small(0.81)}$ & $45.01_{\small(1.20)}$ & $58.26_{\small(0.32)}$ & $77.25_{\small(0.83)}$  & $47.83_{\small(0.68)}$   \\
 & DeepFM     & $51.32_{\small(1.78)}$ 
 &$67.07_{\small(2.86)}$  & $40.15_{\small(1.69)}$ & $51.33_{\small(2.39)}$ & $70.16_{\small(2.93)}$ & $41.15_{\small(2.80)}$ & $52.81_{\small(1.47)}$ &$72.61_{\small(2.42)}$  & $43.12_{\small(1.93)}$   \\
 & AFN        & $59.18_{\small(1.19)}$ & $78.75_{\small(1.98)}$  & $49.19_{\small(1.98)}$ & $60.88_{\small(0.83)}$ & $79.81_{\small(1.82)}$ & $49.92_{\small(1.57)}$ & $64.42_{\small(1.08)}$ & $80.41_{\small(1.71)}$ & $54.07_{\small(0.38)}$   \\
 & GraphRec   & $60.65_{\small(2.79)}$ & $50.73_{\small(3.66)}$ & \underline{$54.77_{\small(3.26)}$} & \cellcolor{LightYellow}$\mathbf{75.49_{\small(2.38)}}$ & $56.05_{\small(3.73)}$ & \cellcolor{LightYellow}$\mathbf{72.20_{\small(2.67)}}$ & \cellcolor{LightYellow}$\mathbf{87.71_{\small(1.72)}}$ & $62.06_{\small(3.56)}$ & \cellcolor{LightYellow}$\mathbf{86.31_{\small(1.89)}}$   \\
\hline           
\end{tabular}
}
\end{table*}}

\section{Experimental Studies}
\label{sec:experiment}

In this section, we give the test setting (\cref{subsec:expsetting}) and report our substantial experimental results in the following facets: 
\ding{172} Compare the effectiveness of \HIRE with the state-of-the-art approaches under 3 cold-start scenarios (\cref{subsec:comresults}). 
\ding{173} Compare the test efficiency of \HIRE with the baseline approaches (\cref{subsec:efficiency}).
\ding{174} Investigate the sensitivity of \CT regarding key hyper-parameter configurations (\cref{subsec:sensitiveanalysis}). 
\ding{175} Study the influence of different attention layers via an ablation study (\cref{subsec:ablationstudy}).
\ding{176} Conduct a case study for HIRE on a movie rating task. (\cref{subsec:casestudy}).

\comment{
We discuss experimental setup in Section \ref{subsec:expsetting} and show effectiveness of \CT with the baseline under different cold start scenarios in the following Section \ref{subsec:comresults}. Sensitive analysis is followed in Section~\ref{subsec:sensitiveanalysis} and Section~\ref{subsec:ablationstudy} conducts ablation study on \CT regarding attention layers. A case study is included at last in~\ref{subsec:casestudy}.
}

\subsection{Experimental Setup}
\label{subsec:expsetting}
\stitle{Datasets:}
{We use 3 widely used datasets to evaluate \HIRE and the baseline approaches, whose profiles are summarized in Table~\ref{tab:dataset}.} 
\shadd{\Movielens \cite{movielens} contains the ratings of users for movies, and the users and movies are associated with rich attributes.
Following the previous following the previous literature~\cite{MeLU}, We randomly split 80\% of users for training and validation, and 20\% for test. 
movies released before 1997 are used for training and validation while the remaining are used for test, resulting in an approximate ratio of 8:2.
\Bookcrossing \cite{bookcrossing} is collected from Book-Crossing, containing ratings of users on books. 
\Douban \cite{douban1, douban2} contains the ratings of users on musics, which are extracted from a rating website Douban. 
This dataset also contains friendship relations between users.
Since users/items do not have attributes, we use the embeddings of the user/item ID as the attributes, which are generated by learnable linear weights.
For \Bookcrossing and \Douban, we randomly split $70\%$ of users/items into the train and validation set and $30\%$ into the test set, respectively.
For all the datasets, train and validate sets are further split randomly by a ratio of 7:1.
The test users/items are regarded as cold users/items where
10\% of the interactions, i.e., no more than 3 interactions, in the  datasets can be
exploited by the recommendation models and the remaining
90\% are used for model evaluation. 
}

\comment{
We use 3 widely used public available recommendation datasets (\Movielens, \Douban and \Bookcrossing) to evaluate \CT. Table \ref{tab:dataset} summarizes the details of datasets.
\begin{itemize}[noitemsep,topsep=0pt,parsep=5pt,partopsep=0pt,leftmargin=*]
    \item \Movielens \cite{movielens} is a dataset including $1,000,209$ ratings from $6,040$ users for $3,706$ movies. The ratings range from 1 to 5. The user attributes have gender, age, occupation and zip code. The item attributes include publication year, rate, gender and actor. We kept users whose item-consumption history length is between $13$ and $100$. We randomly divided the users into existing users and new users in an approximate ratio of 8:2. We divided movies by publication year 1997. For movies released before 1997 we regarded as existing items and for those released after 1997 we regarded as new items.
    \item \MovieTweetings \cite{movietweetings} is collected from Twitter on a daily basis. It contains $71,687$ users, $38,011$ items and $921,117$ ratings. The ratings range from 1 to 10. The user has no attribute and item contains genre as contents. Users are randomly divided into existing users and new users with a ratio of 8:2. We divided the movies into $80\%$ exisiting movies and $20\%$ new movies by publication year 2016.
    \item \Bookcrossing \cite{bookcrossing} is collected from Book-Crossing, containing $278,858$ users, $271,379$ books and $1,149,780$ ratings. The ratings range from 1 to 10. Some users or items without attributes are filtered. The user attributes include age and the item attributes include publication year. We divide both users and items in a ratio of 7:3. We selected seventy percent of the users as existing users and remaining as new users. For items, we selected seventy percent of books as existing books and remaining as new books.
\end{itemize} 
}

\stitle{Baselines:} 
\shadd{To comprehensively evaluate HIRE, we compare with the following 12 baselines which fell into 6 categories:}
\comment{
i.e., (1) a basic strategy, \Popularity, (3) 4 neural CF-based approaches, \NEUMF~\cite{ncfgithub}, \DEEPFM~\cite{deepfmgithub}, \WIDE~\cite{deepfmgithub} and \AFN~\cite{deepfmgithub}, (4) a social recommendation approach, \GraphRec~\cite{graphrecgithub}, (5) 2 HIN-based approaches, \MetaHIN~\cite{metahingithub} and \GraphHINGE~\cite{hingegithub}, (6) 3 meta-learning approaches for cold-start recommendation, \MAMO~\cite{mamogithub}, \TANP~\cite{tanpgithub} and \MELU~\cite{melugithub}.
}
\ding{182} \shadd{Basic Strategy: \Popularity predicts ratings by summing the number of interactions and normalizing for each item.}
\ding{183} CF-based baselines:
\NEUMF~\cite{NCF}, \WIDE~\cite{WIDE}, \DEEPFM~\cite{deepFM} and \AFN~\cite{AFN} adopt different neural network layers to model the interactions from user-item pairs. 
%
%
\ding{184} Social recommendation:
\GraphRec~\cite{fan2019graph} models the users and items respectively by aggregating their local interactions via Graph Neural Network. Aggregation of the social relation among users is used to enhance user embeddings. 
We only compare \GraphRec on \Douban where the social relationship between users is available.
\ding{185} HIN-based baselines:
\GraphHINGE~\cite{jin2020efficient} aggregates rich interactive patterns from an HIN. 
{
\MetaHIN~\cite{metahin} incorporates multifaceted semantic contexts induced by meta-paths.
We test \MetaHIN and \GraphHINGE only on \Movielens which provides sufficient attributes for constructing an HIN. 
Specifically, we define user, movie, genre, occupation, and age as different node types and create links between nodes, such as movie-genre interaction, user-age interaction, and user-movie interaction. 
For the other two datasets \Bookcrossing and \Douban, due to lack of sufficient attributes of user/item, the two baselines cannot be applied to them.
}
\ding{186} Meta-learning baselines: \MAMO~\cite{MAMO}, \MELU~\cite{MeLU} and \TANP~\cite{TaNP} adopt meta-learning for cold-start recommendation in a multi-task fashion. \MAMO and \MELU support all the 3 cold-start scenarios while \TANP only supports user cold-start originally.
Here, we extend the task settings of \TANP to item cold-start and user \& item cold-start.
%
\ding{187} \shadd{Transformer-based baselines: \transgnn~\cite {zhang2024transgnn} integrates transformer and GNN layers to mutually enhance their capabilities.}

\comment{
\begin{itemize}[noitemsep,topsep=0pt,parsep=5pt,partopsep=0pt,leftmargin=*]

\item  
\item {\GraphRec \cite{fan2019graph}. \GraphRec can jointly model social relation and rating prediction by capturing interactions and opinions in the user-item graph. Since it needs social relations of users, we only compare this method in \Douban dataset.}

\item {\GraphHINGE \cite{jin2020efficient}. To capture the interactive patterns through meta-path neighborhoods, GraphHINGE was proposed for learning interactions in HINs. We take \Movielens to construct HIN since it contains adequate feature information for users or items.}

\item Task-adaptive Neural Process \cite{TaNP} (\TANP).
Task-adaptive neural process belongs to a member of neural process family. \TANP maps observed interactions to predictive distribution and introduce a task-adaptive mechanism to learn the relevance of different tasks as well as modulating the decoder parameters.

\item Meta-Learned User Preference Estimator \cite{MeLU} (\MELU).
\MELU proposed a MAML-based recommender system to solve cold-start problems. It can estimate user preferences based on only a few consumed items and then rapidly adopt new task.

\comment{\item Memory-Augmented Meta-Optimization \cite{MAMO} (\MAMO).
\MAMO is a memory-augmented framework of MAML. Specifically, they introduce feature-specific memories to guide the model with personalized parameter initialization and task-specific memories to quickly predict the user preference.} 
\end{itemize}}

\stitle{Implementation details:}
We introduce the model and training details of \HIRE as well as the baselines.
For \HIRE, the model is equipped with $3$ HIM blocks in which each MHSA layer has $8$ heads and the hidden dimension of each head is set to $16$. The model is trained by a LAMB optimizer \cite{LAMB} with $\beta=(0.9,0.999)$ and $\epsilon=10^{-6}$, and a Lookahead \cite{lookahead} wrapper with slow update rate $\alpha=0.5$ and $k = 6$ steps between updates. 
We use a flat-then-anneal learning rate scheduler which flats at the base learning rate for $70\%$ of steps, and then anneals following a cosine schedule to $0$ by the end of training. 
We set the base learning rate to $10^{-3}$ and the gradient clip threshold to $1.0$. The optimizer and the scheduler are widely used for training language models composed of deep MHSA layers. 
Regarding the prediction contexts, we set the number of users and items in each context to 32 by default. In the training and test, $90\%$ of observed ratings are masked for prediction and the remaining $10\%$ are associated with the context input. The learning framework of \CT is built on PyTorch. 

For the baseline approaches, we use their released source code and keep the hyper-parameters as their default settings. All the models are trained until convergence.
For the four CF-based approaches, \GraphRec and \GraphHINGE, to achieve a fair comparison, we use all the observed user-item ratings in the sampled training contexts, together with the $10\%$ unmasked user-item ratings in the test context as the ground truth to train the models.
\shadd{We use the prediction context to test \transgnn, where a few number of user-item interaction is used for training and the remaining is used for test.} 
For meta-learning approaches, in each task, $10\%$ of the observed ratings are used as the ground truth in the support set and the remaining $90\%$ are used in the query set.
Training and testing are conducted on one Tesla V100 with 16GB memory.
\comment{
The source of all baselines have been released and we modify the data input, parameters and evaluations to use the same settings as \HIRE for a fair comparison.
For \TANP, \MELU and \HIRE, they consist several linear layers which embed different features of users/items. The input dimension is original dimension of features while the output dimension is fixed as $16$. Other parameter settings of baselines are their default.
 For \HIRE, we stacked $3$ \HIM blocks, and set $8$ heads for each multi-head self-attention layer. \HIRE is trained with LAMB optimizer \cite{LAMB} ($\beta=(0.9,0.999)$, $\epsilon=1e-6$) and a Lookahead \cite{lookahead} wrapper with slow update rate $\alpha=0.5$ and $k=6$ steps between updates. A flat-then-anneal learning rate schedule with cosine decay is adopted. It flats at base learning rate for $0.7$ of steps, and then anneals following a cosine schedule to $0$ by the end of training. Base learning rate is $1e-3$ and gradient clip is at $1$. The MLP layer maps outputs from \HIM to dimension $1$.
}


\comment{
 CF-based: 
To satisfy the fairness of comparison, we use the user-item pair in training tasks $\mathcal{D}=\{\mathcal{T}_j\}_{j=1}^T$ and those without masks in test tasks for training and then transfer the trained model to those masked pairs in test tasks $\mathcal{T}^*$ to test. There are $1024$ training tasks $10\%$ ratings without masks in tasks. The number of test tasks is depend on the number of users in test set. All data used in these baselines are the same as \HIRE.

Social:  We follow the settings in CF-based baselines. Furthermore, we add social relation of test users in test tasks.

HIN: \GraphHINGE follows the configuration of CF-based baselines since it cannot directly apply to cold-start scenarios. For each task, we can construct an HIN to capture heterogeneous interaction information.

Meta: 
Meta-model are trained with $1024$ training tasks and evaluated in the same number of test tasks in CF-based baseline. A task has to be divided into support set and query set and only query set of test tasks are predicted for recommendation. We set the ratio of support set as $0.1$ for fairly comparing. 

}

\stitle{Evaluation Metrics:}
{We adopt 3 commonly used metrics following the previous work~\cite{TaNP}} to evaluate the recommendation performance in our testing, including: Precision, Mean Average Precision (MAP) and Normalized Discounted Cumulative Gain \cite{jarvelin2002cumulated} (NDCG).
Top $k$ actual rating values sorted by predicted rating values are used to calculate the above metrics, where $k \in \{5,7,10\}$. 

\comment{
\shadd{Three frequently-used metrics to evaluate recommendation performance in the study are as follows: Precision (\PRE), Mean Average Precision (\MAP) and Normalized Discounted Cumulative Gain \cite{jarvelin2002cumulated} (\NDCG) as shown in Eq.~\ref{eq:pre}-\ref{eq:dcg}. 
\begin{align}
    &Pre_k=\frac{|\omega_u \cap \hat{\omega_u}(1:k)|}{k},\label{eq:pre}\\
    &AP^u_k=\frac{1}{|\omega_u|}\sum_{i\in \omega_u}\frac{\sum_{j\in \omega_u}h(p_{j}<p_{i})+1}{p_i},\label{eq:ap}\\
    &MAP=\frac{\sum_{u\in \mathcal{U}}AP_u}{|\mathcal{U}|},\label{eq:map}\\
    &NDCG_k=\frac{1}{|\mathcal{U}|}\sum_{u\in \mathcal{U}}\frac{DCG^u_k}{IDCG^u_k}, \label{eq:ndcg1}\\
    &DCG^u_k=\sum_{i=1}^{k}\frac{r_{ui}}{\log_2(1+i)},\label{eq:dcg}
\end{align}
where $\omega_u$ and $\hat{\omega_u}(1:k)$ are ground truth ranking list and list of top $k$ predicted items for user $u$. $p$ represents the predicted ranking position, $p_{j}<p_{i}$ means item $j$ ranked before item $i$ and $h(\cdot)$ is a function indicating the cumulative count of number in ground truth rankings.
$r_{ui}$ is the ground truth rating of user $u$ for the predicted $i$-th ranked item, $\mathcal{U}$ is the set of users to be predicted. $IDCG^u_k$ refers to the best possible $DCG_u^k$.
Top $k$ actual rating values sorted by predicting rating values are used to calculate the above metrics, where $k=5,7,10$. 
}
}

\begin{table*}[t]
\footnotesize 
\centering
\caption{\shadd{Overall Performance in 3 Cold-Start Scenarios on \Douban (\%)}}
\label{tab:result:douban}
\resizebox{1.0\textwidth}{!}{
\begin{tabular}{c|c|ccc|ccc|ccc}
\toprule
\multirow{2}{*}{Scenarios}                & \multirow{2}{*}{Methods} & \multicolumn{3}{c|}{TOP5}                                           & \multicolumn{3}{c|}{TOP7}                                           & \multicolumn{3}{c}{TOP10}                                          \\ 
 &                          & Precision            & NDCG                 & MAP                  & Precision            & NDCG                 & MAP                  & Precision            & NDCG                 & MAP                  \\ \midrule 
\multirow{11}{*}{\UC}      
 & \shadd{\Popularity}       &\shadd{{$51.43_{\small (0.25)}$}}& 
\shadd{{$86.58_{\small (0.06)}$}}  & \shadd{{$40.01_{\small (0.29)}$}} & \shadd{$56.03_{\small (0.27)}$} & \shadd{{$87.24_{\small (0.06)}$}} & \shadd{$43.12_{\small (0.34)}$} & \shadd{$64.46_{\small (0.23)}$} &\shadd{{$88.52_{\small (0.06)}$}}  & \shadd{$51.20_{\small (0.37)}$}                \\
 & \shadd{\transgnn}                & 
\shadd{$38.54_{\small (0.90)}$} & \shadd{$38.45_{\small (1.35)}$}  & \shadd{$21.76_{\small (1.24)}$} & \shadd{$38.69_{\small (2.10)}$} & \shadd{$38.51_{\small (1.53)}$} & \shadd{$23.55_{\small (1.51)}$} & \shadd{$38.96_{\small (1.30)}$} &\shadd{$38.64_{\small (1.80)}$}  & \shadd{$25.60_{\small (2.10)}$}                \\
& NeuMF      & $44.43_{\small(1.60)}$  & $33.34_{\small(1.99)}$ & $40.56_{\small(1.64)}$ & $58.72_{\small(1.24)}$ & $34.71_{\small(1.91)}$ & $56.23_{\small(1.30)}$ & $72.61_{\small(0.71)}$ & $37.65_{\small(1.68)}$ & $71.13_{\small(0.80)}$    \\
 & Wide\&Deep & $54.42_{\small(1.05)}$ & $73.76_{\small(0.70)}$ & $44.43_{\small(1.75)}$ & $55.66_{\small(0.31)}$ & $75.95_{\small(0.81)}$ & $45.01_{\small(1.20)}$ & $58.26_{\small(0.32)}$ & $77.25_{\small(0.83)}$  & $47.83_{\small(0.68)}$   \\
 & DeepFM     & $51.32_{\small(1.78)}$ 
 &$67.07_{\small(2.86)}$  & $40.15_{\small(1.69)}$ & $51.33_{\small(2.39)}$ & $70.16_{\small(2.93)}$ & $41.15_{\small(2.80)}$ & $52.81_{\small(1.47)}$ &$72.61_{\small(2.42)}$  & $43.12_{\small(1.93)}$   \\
 & AFN        & $59.18_{\small(1.19)}$ & $78.75_{\small(1.98)}$  & $49.19_{\small(1.98)}$ & $60.88_{\small(0.83)}$ & $79.81_{\small(1.82)}$ & $49.92_{\small(1.57)}$ & $64.42_{\small(1.08)}$ & $80.41_{\small(1.71)}$ & $54.07_{\small(0.38)}$   \\
 & GraphRec   & $60.65_{\small(2.79)}$ & $50.73_{\small(3.66)}$ & \underline{$54.77_{\small(3.26)}$} & \cellcolor{LightYellow}$\mathbf{75.49_{\small(2.38)}}$ & $56.05_{\small(3.73)}$ & \cellcolor{LightYellow}$\mathbf{72.20_{\small(2.67)}}$ & \cellcolor{LightYellow}$\mathbf{87.71_{\small(1.72)}}$ & $62.06_{\small(3.56)}$ & \cellcolor{LightYellow}$\mathbf{86.31_{\small(1.89)}}$   \\
 & MAMO        & $60.98_{\small(0.07)}$ & $73.56_{\small(0.05)}$ & $51.01_{\small(0.05)}$ & $67.96_{\small(0.05)}$ & $75.94_{\small(0.02)}$ & $57.41_{\small(0.04)}$ & $80.06_{\small(0.03)}$ & $80.60_{\small(0.02)}$ & $72.09_{\small(0.04)}$ \\
 & TaNP       & \underline{$64.08_{\small(3.40)}$}  
 & \underline{$90.20_{\small(1.04)}$} & $54.65_{\small(4.47)}$ & $69.06_{\small(3.08)}$ & \underline{$91.00_{\small(0.94)}$} & $59.19_{\small(4.22)}$ & $78.24_{\small(3.86)}$ & \underline{$92.59_{\small(1.04)}$} & $70.44_{\small(5.30)}$    \\
 & MeLU       & $45.42_{\small(4.80)}$  & $64.52_{\small(1.81)}$ & $34.63_{\small(4.57)}$ & $50.54_{\small(7.06)}$ & $66.23_{\small(2.69)}$ & $38.15_{\small(6.93)}$ & $59.53_{\small(7.98)}$ & $69.30_{\small(3.52)}$ & $47.23_{\small(8.85)}$   \\
 & HIRE       & \cellcolor{LightYellow}$\mathbf{71.52_{\small(3.91)}}$ & \cellcolor{LightYellow}$\mathbf{92.69_{\small(1.31)}}$ & \cellcolor{LightYellow}$\mathbf{65.95_{\small(5.74)}}$ & \underline{$74.59_{\small(2.48)}$} & \cellcolor{LightYellow}$\mathbf{93.28_{\small(1.22)}}$ & \underline{$68.37_{\small(4.20)}$} & \underline{$80.46_{\small(1.30)}$} & \cellcolor{LightYellow}$\mathbf{94.79_{\small(1.11)}}$ & \underline{$75.21_{\small(1.08)}$} 
\\ \midrule 
\multirow{11}{*}{\IC}
 & \shadd{\Popularity}       &\shadd{$47.35_{\small (0.40)}$}& 
\shadd{{$84.35_{\small (0.22)}$}}  & \shadd{$36.10_{\small (0.45)}$} & \shadd{$54.63_{\small (0.08)}$} & \shadd{{$85.44_{\small (0.12)}$}} & \shadd{$42.06_{\small (0.30)}$} & \shadd{$64.85_{\small (0.09)}$} &\shadd{{$87.12_{\small (0.09)}$}}  & \shadd{$51.76_{\small (0.14)}$}                \\
 & \shadd{\transgnn}                & 
\shadd{$54.67_{\small (2.30)}$} & \shadd{$50.73_{\small (2.01)}$}  & \shadd{$44.49_{\small (0.84)}$} & \shadd{$59.42_{\small (0.99)}$} & \shadd{$55.10_{\small (0.59)}$} & \shadd{$47.38_{\small (0.37)}$} & \shadd{$61.33_{\small (1.15)}$} &\shadd{$57.51_{\small (0.47)}$}  & \shadd{$49.39_{\small (0.25)}$}                \\
& NeuMF      & $39.19_{\small(4.42)}$ & $43.05_{\small(4.28)}$ & $30.50_{\small(1.83)}$  & $46.88_{\small(1.67)}$ & $44.23_{\small(4.13)}$ & $36.93_{\small(2.12)}$ & $56.04_{\small(0.88)}$ & $46.44_{\small(3.95)}$ & $45.36_{\small(3.09)}$ \\
 & Wide\&Deep & $22.85_{\small(0.80)}$  & $44.27_{\small(1.78)}$ & $17.87_{\small(0.59)}$ & $25.00_{\small(1.53)}$ & $44.53_{\small(1.70)}$ & $18.78_{\small(1.27)}$ & $29.37_{\small(1.30)}$ & $45.33_{\small(1.72)}$ & $21.68_{\small(1.05)}$ \\
 & DeepFM     & $23.90_{\small(0.23)}$  & $46.33_{\small(0.21)}$ & $18.56_{\small(0.29)}$ & $25.74_{\small(0.19)}$ & $46.65_{\small(1.51)}$ & $18.81_{\small(0.53)}$ & $29.70_{\small(0.68)}$ & $47.31_{\small(1.48)}$ & $21.30_{\small(0.74)}$  \\
 & AFN        & $26.00_{\small(2.71)}$   & $49.15_{\small(2.67)}$ & $20.44_{\small(2.17)}$ & $28.75_{\small(2.10)}$ & $49.93_{\small(2.38)}$ & $21.60_{\small(1.75)}$ & $32.98_{\small(2.59)}$ & $50.56_{\small(2.65)}$ & $24.05_{\small(1.83)}$ \\
 & GraphRec   & $34.60_{\small(2.98)}$  & $39.73_{\small(3.60)}$ & $28.47_{\small(3.93)}$ & $43.28_{\small(3.28)}$ & $42.73_{\small(4.09)}$ & $35.97_{\small(4.36)}$ & $53.88_{\small(2.70)}$ & $46.39_{\small(4.24)}$ & $45.08_{\small(4.69)}$  \\
 & MAMO       & \underline{$59.80_{\small(0.11)}$}  & $72.50_{\small(0.05)}$ & \underline{$49.86_{\small(0.10)}$}  & \cellcolor{LightYellow}$\mathbf{66.96_{\small(0.11)}}$ & $74.95_{\small(0.04)}$ & \underline{$56.39_{\small(0.12)}$} & \cellcolor{LightYellow}$\mathbf{79.43_{\small(0.06)}}$ & {$79.78_{\small(0.03)}$} & \cellcolor{LightYellow}$\mathbf{71.20_{\small(0.06)}}$  \\
 & TaNP       & $49.45_{\small(3.03)}$ & \underline{$85.02_{\small(0.90)}$} & $38.08_{\small(3.26)}$ & $54.79_{\small(3.96)}$ & \underline{$85.90_{\small(1.00)}$} & $42.30_{\small(4.62)}$ & $63.28_{\small(5.95)}$ & \underline{$87.40_{\small(1.35)}$} & $51.07_{\small(7.50)}$  \\
 & MeLU       &$50.87_{\small(3.63)}$ & $66.50_{\small(2.21)}$  & $38.76_{\small(3.35)}$ & $56.19_{\small(6.85)}$ & $68.59_{\small(3.71)}$ & $43.54_{\small(8.22)}$ & $65.12_{\small(9.02)}$ & $71.79_{\small(4.93)}$ & $57.70_{\small(5.76)}$ 
 \\
 & HIRE       & \cellcolor{LightYellow}$\mathbf{61.28_{\small(4.67)}}$ & \cellcolor{LightYellow}$\mathbf{89.26_{\small(1.59)}}$ & \cellcolor{LightYellow}$\mathbf{54.96_{\small(6.79)}}$ & \underline{$64.63_{\small(3.07)}$} & \cellcolor{LightYellow}$\mathbf{89.83_{\small(1.47)}}$ & \cellcolor{LightYellow}$\mathbf{57.08_{\small(5.13)}}$ & \underline{$70.66_{\small(0.67)}$} & \cellcolor{LightYellow}$\mathbf{91.22_{\small(1.36)}}$ & \underline{$63.05_{\small(1.96)}$} \\\midrule 
\multirow{11}{*}{\UIC}  
 & \shadd{\Popularity}       &\shadd{${47.54_{\small (0.39)}}$}& 
\shadd{\underline{$87.82_{\small (0.19)}$}}  & \shadd{${42.24_{\small (0.57)}}$} & \shadd{$54.10_{\small (0.26)}$} & \shadd{{$87.95_{\small (0.05)}$}} & \shadd{${46.47_{\small (0.27)}}$} & \shadd{$70.82_{\small (0.05)}$} &\shadd{${90.14_{\small (0.06)}}$}  & \shadd{$62.91_{\small (1.12)}$}                \\
 & \shadd{\transgnn}                & 
\shadd{$27.33_{\small (1.15)}$} & \shadd{$30.34_{\small (1.50)}$}  & \shadd{$25.07_{\small (0.71)}$} & \shadd{$30.47_{\small (1.65)}$} & \shadd{$32.91_{\small (1.88)}$} & \shadd{$27.33_{\small (1.01)}$} & \shadd{$30.67_{\small (2.30)}$} &\shadd{$33.62_{\small (2.26)}$}  & \shadd{$29.11_{\small (1.02)}$}                \\
& NeuMF      & $27.63_{\small(0.55)}$ & $35.97_{\small(1.91)}$  & $22.02_{\small(0.63)}$ & $28.57_{\small(1.30)}$  & $36.16_{\small(1.18)}$ & $22.37_{\small(0.99)}$ & $33.40_{\small(2.35)}$ & $38.98_{\small(0.98)}$ & $24.78_{\small(0.79)}$ \\
 & Wide\&Deep & $8.42_{\small(2.72)}$   &$13.03_{\small(1.53)}$  & $7.30_{\small(1.67)}$ & $8.48_{\small(2.56)}$ & $14.35_{\small(1.18)}$ & $7.31_{\small(1.44)}$ & $8.57_{\small(2.02)}$ & $16.15_{\small(1.43)}$ & $8.32_{\small(1.77)}$ \\
 & DeepFM     & $5.78_{\small(3.82)}$ & $11.03_{\small(6.67)}$ & $4.82_{\small(2.83)}$ & $5.96_{\small(3.30)}$ & $12.48_{\small(6.63)}$ & $5.21_{\small(2.92)}$ & $7.52_{\small(0.36)}$ & $14.50_{\small(1.82)}$ & $6.12_{\small(0.84)}$ \\
 & AFN        & $5.26_{\small(0.98)}$ & $11.23_{\small(1.82)}$ & $4.76_{\small(0.26)}$ & $5.88_{\small(1.67)}$ & $12.86_{\small(1.92)}$ & $5.27_{\small(1.26)}$ & $6.39_{\small(1.00)}$ &$14.84_{\small(1.95)}$  & $5.52_{\small(1.15)}$ \\
 & GraphRec   & $35.68_{\small(2.83)}$ 
& $39.00_{\small(1.68)}$ & $26.24_{\small(.01.78)}$ & $40.66_{\small(2.71)}$ & $40.43_{\small(1.65)}$ & $28.85_{\small(2.34)}$ & $47.00_{\small(1.81)}$ & $42.76_{\small(1.28)}$ & $33.26_{\small(1.49)}$ \\
 & MAMO    & $60.09_{\small(0.12)}$ & $72.78_{\small(0.04)}$ & \underline{$50.37_{\small(0.11)}$} & $67.50_{\small(0.16)}$ & $75.31_{\small(0.05)}$ & $57.21_{\small(0.16)}$ & \cellcolor{LightYellow}$\mathbf{79.76_{\small(0.06)}}$& $80.07_{\small(0.03)}$ & \cellcolor{LightYellow}$\mathbf{71.73_{\small(0.08)}}$ \\    
 & TaNP       & \underline{$60.67_{\small(2.44)}$} & $87.34_{\small(0.40)}$ & $49.82_{\small(2.17)}$ & \cellcolor{LightYellow}$\mathbf{69.62_{\small(0.89)}}$ & \underline{$89.03_{\small(0.30)}$} & \underline{$58.05_{\small(1.66)}$} & $77.55_{\small(1.74)}$ & \underline{$91.00_{\small(0.39)}$} & $66.90_{\small(1.71)}$  \\
 & MeLU       & $50.32_{\small(1.73)}$ & $67.37_{\small(0.70)}$ & $39.34_{\small(1.11)}$ & $55.44_{\small(2.73)}$ & $68.75_{\small(1.16)}$ & $43.55_{\small(2.28)}$ & $67.17_{\small(1.25)}$ & $72.94_{\small(0.78)}$ & $56.26_{\small(1.36)}$ \\
 & HIRE       &\cellcolor{LightYellow}{$\mathbf{62.66_{\small(1.95)}}$} & \cellcolor{LightYellow}$\mathbf{89.02_{\small(2.14)}}$ & \cellcolor{LightYellow}{$\mathbf{54.16_{\small(5.86)}}$} & \underline{$69.53_{\small(1.69)}$} & \cellcolor{LightYellow}$\mathbf{90.17_{\small(1.94)}}$ & \cellcolor{LightYellow}{$\mathbf{61.48_{\small(3.24)}}$} & \underline{$78.33_{\small(2.51)}$} & \cellcolor{LightYellow}{$\mathbf{92.47_{\small(1.66)}}$} & \underline{$71.06_{\small(2.00)}$}   \\ \bottomrule 
\end{tabular}
}
\end{table*}

\subsection{Comparison Results}
\label{subsec:comresults}


We investigate the overall performance of \HIRE in 3 cold-start scenarios, i.e., user cold-start (\UC), item cold-start (\IC), and user \& item cold-start (\UIC), in comparison with the 12 baseline approaches. We also investigate whether \HIRE can effectively handle regular cases by using data from warm states and we compare \HIRE against several competitive baselines. This means that the test users and items may also appear in the training dataset and have a substantial history of interactions. The results are reported in Table~\ref{tab:result:warmstate}.
Table~\ref{tab:result:movielens}, \ref{tab:result:bookcross} and \ref{tab:result:douban} list the test results on 3 datasets, respectively, where the first-best (yellow) and the second-best (underlined) scores are highlighted. 
In general, \HIRE achieves the best test accuracy in most cases. 
The Precision, NDCG and MAP of \HIRE succeed all the baselines {0.21, 0.29, 0.22} on average, respectively. The superiority of \HIRE is reflected in improving NDCG while keeping high Precision and MAP. 

In Table~\ref{tab:result:warmstate}, \HIRE outperforms these baselines, demonstrating its capability in handling regular warm cases effectively. Furthermore, since in regular cases, the prediction context involves a substantial amount of historical interactions for the test users/items, \HIRE achieves better performance in regular cases than in the cold-start scenarios in most cases. 

\shadd{\Popularity achieves a very competitive performance on \Movielens and \Douban, since the correlation  between the number of interactions and the rating is positive in these datasets. \transgnn does not perform well in cold-start scenarios since it relies heavily on user-item interactions.}
In contrast to neural CF-based approaches, \NEUMF, \WIDE, \DEEPFM and \AFN, \HIRE outperforms them on all the datasets significantly. The CF-based approaches only model a single type of user-item interaction via observed ratings, which is difficult to generalize to new users/items with only a few interactions. This reflects modeling heterogeneous interactions is beneficial for cold-start prediction.
{The meta-learning approaches, \MAMO, \TANP and \MELU, outperform the CF-based approaches by a large margin. 
The performance of these 3 approaches dominates the second best result, and is competitive to our \HIRE in some cases.} The meta-learning approaches focus on exploiting parameter sharing and adaption to address cold-start problem. Their complicated adaption module and learning strategy may lead to a difficult learning procedure. 
\HIRE deals with the cold-start problem in a different way that relies on the interaction learned from the data. It concentrates on using a relatively simple neural network framework and learning algorithm to model heterogeneous interactions in a data-driven fashion.


On \Movielens (Table~\ref{tab:result:movielens}), we compared HIN-based baseline \GraphHINGE and \MetaHIN. In the user cold-start scenario, \GraphHINGE gains better performance than CF-based approaches,  indicating that metapath-guided neighborhood can capture the complex and high-order semantics in the HIN. 
{In item cold-start and user \& item cold-start scenarios, \MetaHIN consistently outperforms \GraphHINGE, as indicated by its higher NDCG scores.
However, these two approaches fail to outperform \HIRE and \TANP. It validates again that manually defined heterogeneous patterns may not contribute to estimating preference, while the learned heterogeneous interactions by \HIRE tend to be more reliable.}
We compare the social recommendation baseline, \GraphRec, on \Douban (Table~\ref{tab:result:douban}). \GraphRec performs better in the user cold-start scenario, and can even surpasses \HIRE. 
\GraphRec uses Graph Neural Network to integrate the rating interactions and social relations, which generates better user representations. However, we find that \GraphRec is less effective in scenarios with cold items, where the social relations between users are less helpful for cold items. 

\begin{figure}[t]
	\centering
	\includegraphics[width=0.9\columnwidth]{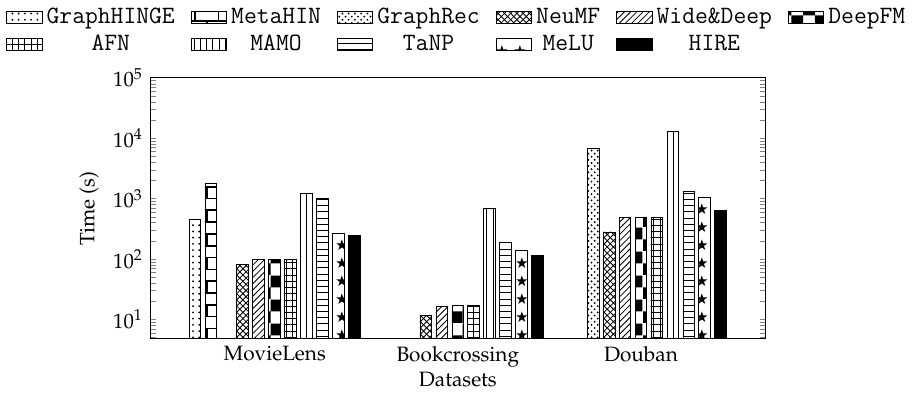}
	\caption{{Total Test Time (in second)}}
        \label{fig:test time}
\end{figure}

\begin{figure}[t]
        \includegraphics[width=0.9\columnwidth]{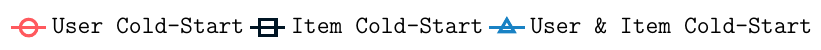}
	\centering
	\begin{tabular}[h]{c}
		\subfigure[\# HIM Blocks] {
		\hspace{-0.4cm}	\includegraphics[width=0.3\columnwidth]{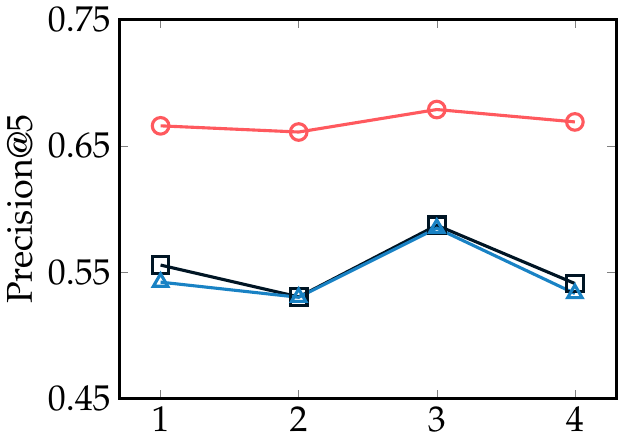}
			\label{fig:layerpre}
		}
   \subfigure[\# HIM Blocks] {
		\hspace{-0.4cm}		\includegraphics[width=0.3\columnwidth]{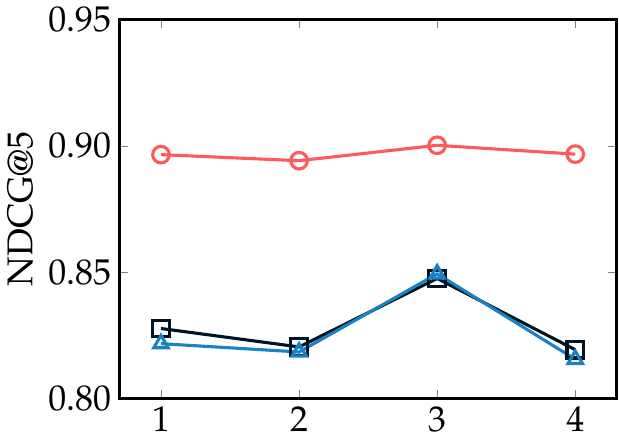}
			\label{fig:layerndcg}
		} 
	\subfigure[\# HIM Blocks] {
		\hspace{-0.4cm}		\includegraphics[width=0.3\columnwidth]{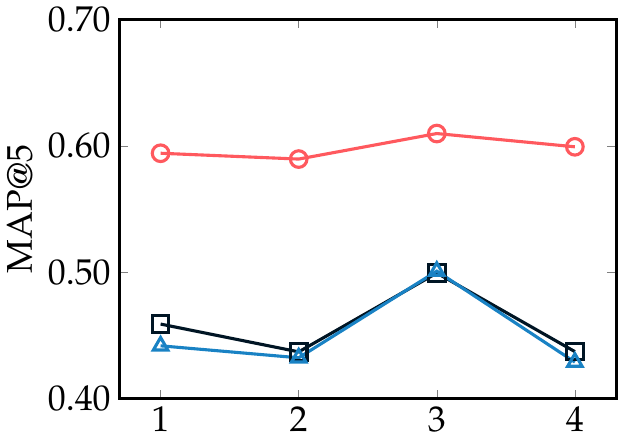}
			\label{fig:layermap}
            }\\
		\subfigure[\# Users/Items] {
			\hspace{-0.4cm} \includegraphics[width=0.3\columnwidth]{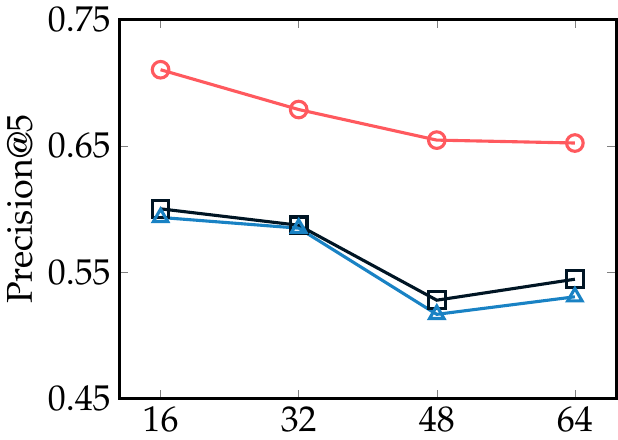}
			\label{fig:rowbookpre}
            }
        \subfigure[\# Users/Items] {
			\hspace{-0.4cm} \includegraphics[width=0.3\columnwidth]{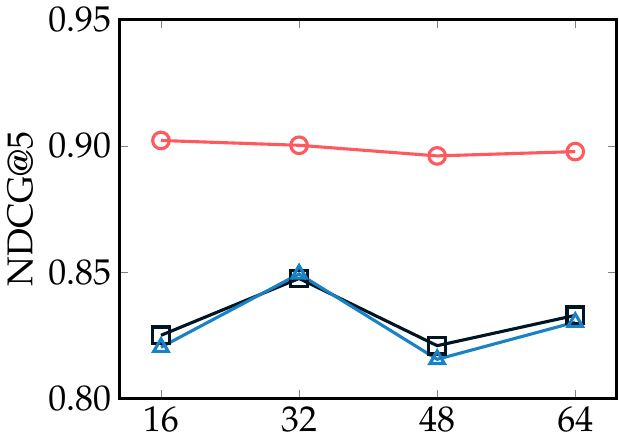}
			\label{fig:rowbookndcg}
		}
             \subfigure[\# Users/Items] {
			\hspace{-0.4cm} \includegraphics[width=0.3\columnwidth]{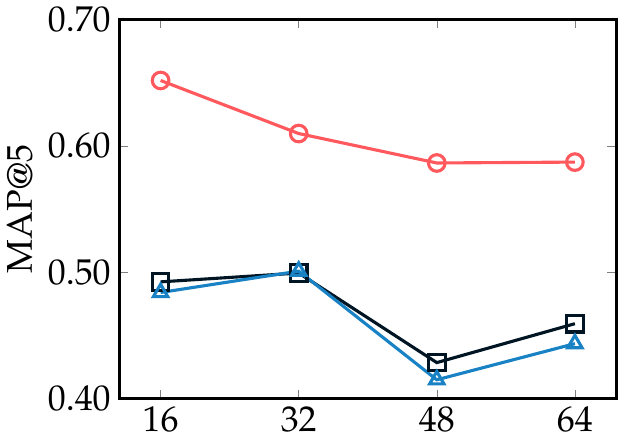}
			\label{fig:rowbookmap}
		} 
	\end{tabular}
	\caption{{Sensitivity Analysis on \Movielens}}
        \label{fig:parameter}
\end{figure}

\begin{table*}[t]
\footnotesize
\centering
\vspace{-2ex}
\caption{Ablation Study for Different Attention Layers on \Movielens (\%)}
\label{tab:layer}
\resizebox{1.0\textwidth}{!}{
\begin{tabular}{c|c c c|c c c|c c c}
\toprule
\multirow{2}{*}{Blocks} & \multicolumn{3}{c|}{User Cold-Start}   & \multicolumn{3}{c|}{Item Cold-Start}   & \multicolumn{3}{c}{User \& Item Cold-Start}  \\ 
 & Pre.@5 & NDCG@5 & MAP@5  & Pre.@5 & NDCG@5 & MAP@5  & Pre.@5 & NDCG@5 & MAP@5  \\\midrule
wo/ Item \& Attribute                   & 44.65 & 78.58 & 32.32 & 43.92 & 76.00   & 31.77 & 46.63 & 77.00   & 34.40 \\
wo/ User \& Attribute                   & 65.52 & 89.26 & 58.38 & 52.68 & 81.74 & 43.01 & 52.27 & 81.38 & 42.39 \\
wo/ User \& Item                    & 67.52 &89.86	&60.40 & 51.63 & 81.28 & 42.02 & 50.67 & 80.79 & 40.73 \\
wo/ User                    & 65.90  & 89.25 & 58.85 & 52.72 & 81.16 & 42.23 & 52.39 & 81.11 & 42.13 \\
wo/ Item                  & 44.61 & 78.66 & 32.38 & 44.14 & 76.10  & 31.93 & 46.87 & 77.00   & 34.47  \\
wo/ A                   & 44.77 & 78.65 & 32.42 & 44.13 & 76.11 & 32.00   & 46.71 & 76.99 & 34.42 \\ \midrule
wo/ Attribute                     & \cellcolor{LightYellow}\textbf{69.99} & \cellcolor{LightYellow}\textbf{91.69} & \cellcolor{LightYellow}\textbf{64.54} & \cellcolor{LightYellow}\textbf{59.89} & \cellcolor{LightYellow}\textbf{86.40} & \cellcolor{LightYellow}\textbf{53.04} & \cellcolor{LightYellow}\textbf{60.30} & \cellcolor{LightYellow}\textbf{86.93} & \cellcolor{LightYellow}\textbf{53.62} \\\bottomrule
\end{tabular}}
\vspace{-2ex}
\end{table*}

\subsection{Efficiency}
\label{subsec:efficiency}
\comment{
We compare the efficiency of \HIRE and the baselines regarding the test time. Fig.~\ref{fig:test time} shows GPU test time of our \HIRE and other 8 baselines on three datasets. CF-based methods take the least amount of time, since it only calculates all the unmasked user-item pairs in test tasks. In the meantime, the model is relatively simple and intuitively takes less time. \HIRE takes a longer time than these traditional methods, however, it takes a shorter time compared to meta-learning approach. \GraphHINGE needs to construct HIN, which results in the second longest time in \Movielens. \GraphRec costs most since it devises additional module to aggregate social relations.
In short, \HIRE trades a moderate amount of time for the best performance.}

{We compare the test time of \HIRE with the baseline approaches, where the total test time for the 3 datasets on a single Tesla V100 GPU is presented in Fig.~\ref{fig:test time}.} 
Recall that the HIN-based approach \GraphHINGE is only applicable for \Movielens and the social recommendation approach \GraphRec is only applicable for \Douban.
{Here, we only compare for the user cold-start since the test time for the 3 cold-start scenarios is similar.}
The CF-based approaches, NeuMF, Wide\&Deep, AFN and DeepFM are the most time-efficiency approaches, since they only take a pair of user and item with their potential features as input, and the model architectures are relatively simple. Our approach, \HIRE, spends a longer time for prediction than the CF-based approaches due to the computational cost of multi-layer MHSA. 
{However, the prediction of \HIRE is faster than that of the other 6 approaches on average.
The meta-learning approaches, \MAMO, \MetaHIN, \TANP and \MELU, leverage extra modules for model adaption. \GraphHINGE needs to sample and model the neighborhood interactions from an HIN, which results in the second longest time in \Movielens. \GraphRec spends the second longest time for prediction since it deploys an additional Graph Neural Network to aggregate social relations. \MAMO is significantly slower than \HIRE, with about one order of magnitude. This is because \MAMO is built upon the MAML algorithm and utilizes two specific memory modules, which have high time and space complexities.
In a nutshell, \HIRE outperforms other methods by achieving the best overall effectiveness while maintaining a competitive prediction efficiency.}

\begin{figure}[t]
        \includegraphics[width=0.5\columnwidth]{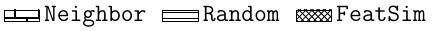}
	\centering
	\begin{tabular}[h]{c}
        \subfigure[Precision] {
		\hspace{-0.4cm}	\includegraphics[width=0.3\columnwidth]{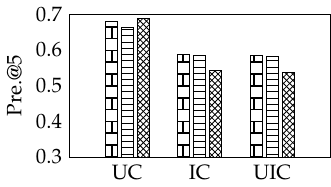}
			\label{fig:samplepre}
		}
        \hspace*{-0.4cm}
		\subfigure[NDCG] {
		\hspace{-0.2cm}	\includegraphics[width=0.3\columnwidth]{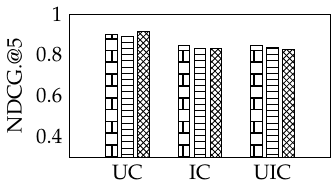}
			\label{fig:samplendcg}
		}
        \hspace*{-0.4cm}
		\subfigure[MAP] {
		\hspace{-0.2cm}	\includegraphics[width=0.3\columnwidth]{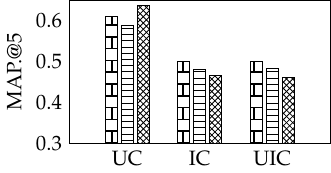}
			\label{fig:samplemap}
            }
	\end{tabular}
	\caption{{Impact of sampling methods on \Movielens}}
        \label{fig:samplemethod}
\end{figure}

\subsection{Sensitivity Analysis}
\label{subsec:sensitiveanalysis}
We investigate the parameter sensitivity of \HIRE with respect to two key hyper-parameter configurations: (1) the number of the Heterogeneous Interaction Modules (HIM) in \HIRE and (2) the number of users/items sampled in one prediction context. 
{Due to space limitation, we present the evaluation results for the metrics at $k=5$ considering it can be regarded as a lower bound for all the metrics.}
First, we train \HIRE model variants by varying the number of HIMs in $\{ 1, 2, 3, 4\}$ on \Movielens, where all the other hyper-parameters are fixed as default.
Fig.~ 
\ref{fig:layerpre}, \ref{fig:layerndcg} and  \ref{fig:layermap} show the test accuracy in the 3 cold-start scenarios. 
Here, \HIRE with $3$ HIMs achieves the best performance, which is consistent to the different cold-start scenarios. As the number of HIMs increases, the model is able to capture increasingly high-order and complex interactions.
However, more HIMs such as 4 will incur the risk of overfiting and lead to performance degradation in practice. 
{
In contrast, for \Bookcrossing and \Douban, in our extra experiments, we observe that 2 and 4 HIM blocks achieve the best performance, respectively. Thereby, different datasets may need different configurations to reach the best result.
}   
Another observation is that \HIRE in user cold-start is less sensitive to the number of HIMs, compared with item and user \& item cold-start. 

In addition, we also compare the accuracy of \HIRE by varying the number of sampled users/items in the training and test context in $\{16, 32, 48, 64\}$, where the results are shown in Fig.~\ref{fig:rowbookpre}, \ref{fig:rowbookndcg} and \ref{fig:rowbookmap}. As the number of users/items increases, the performance of \HIRE may not change monotonically.  More users/items in the context matrices may provide richer information for the prediction, e.g., in the cases of 64 users/items, or bring noise information, e.g., in the cases of 48 users/items, which hurts the performance.
{For \Movielens, setting the number of users/items to 32 yields the most impressive scores, whereas in our extra testing on \Bookcrossing and \Douban, the setting of 64 yields the best scores. We speculate that the reason would be the latter two datasets have less attributes so that they need more users/items in the prediction context.}

\comment{
We further investigate the parameter sensitivity of \HIRE with respect to two main parameters: 1) Different stacking blocks, 2) Number of users/items in a batch. 
\shadd{The experiment is conducted on \Movielens.
Figure~\ref{fig:parameter} shows the performance under different stacking blocks (left) and different number of user/item (right) on \Movielens. \HIRE achieves competitive results when stacking $3$ blocks.
With increasing number of blocks, the model can capture increasingly complex interactions and dependencies. However, too many blocks may incur the risk of overfitting. Thus we set $3$ blocks as default for \HIRE.
As we can see, \HIRE is robust to different blocks in user-cold-start scenario, and it is quite sensitive in other two scenarios.} 

\shadd{The difference between the stacking number of user/item are relatively slight. As the number of user/item increases, it is intuitive to find results deteriorate due to the increasing difficulty of the task. However, the performance of \HIRE  does not monotonically decrease. It may because when number of user/item increase, more implicit reliable interactions can be learned by model, which is helpful to estimate preference. 
We also found that \UC get higher score than other two scenarios and \IC performs better than \UIC a little. The result is consistent with table~\ref{tab:result:movielens}.
}
}

\begin{figure*}[t]
	\centering
	\begin{tabular}[h]{c}
		\subfigure[Weights of \ABU] {
			\includegraphics[width=0.45\columnwidth]{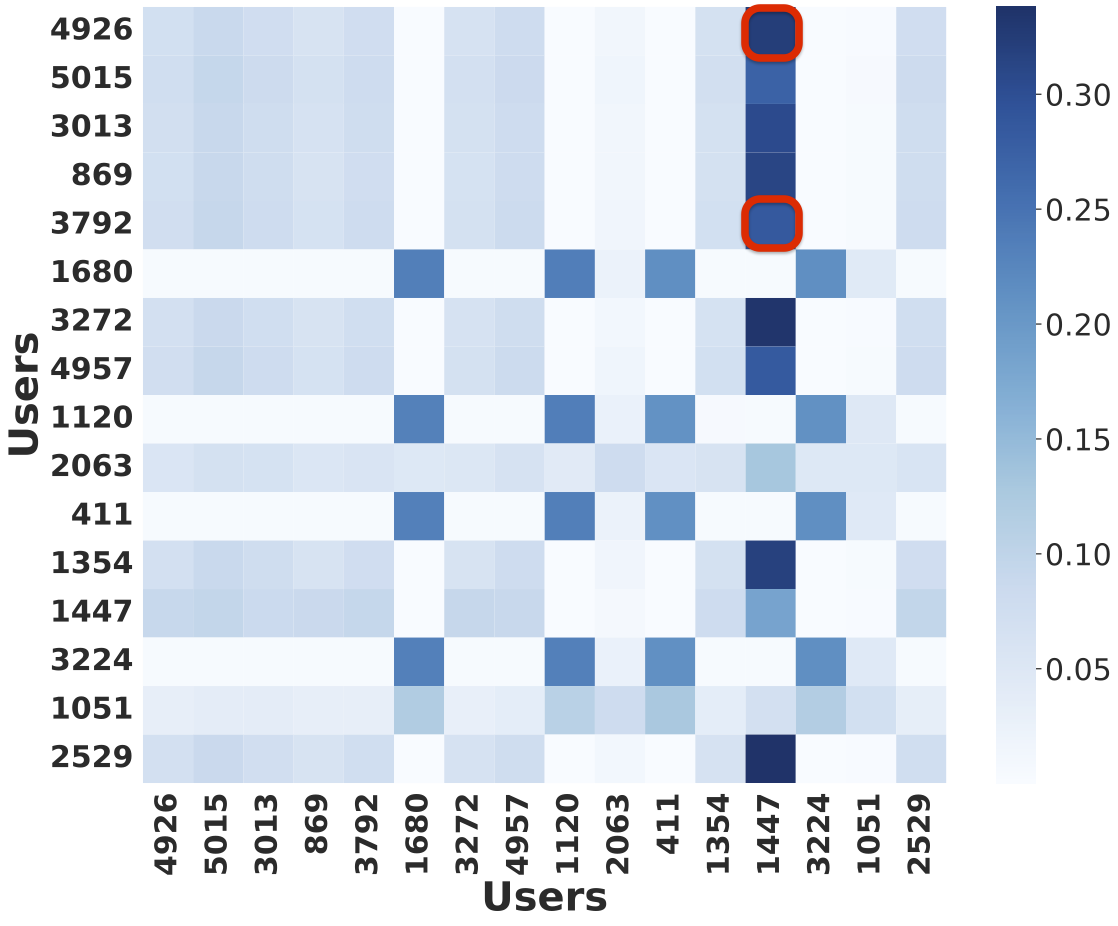}
			\label{fig:userweight}
		}
		\subfigure[Weights of \ABI] {
			\includegraphics[width=0.45\columnwidth]{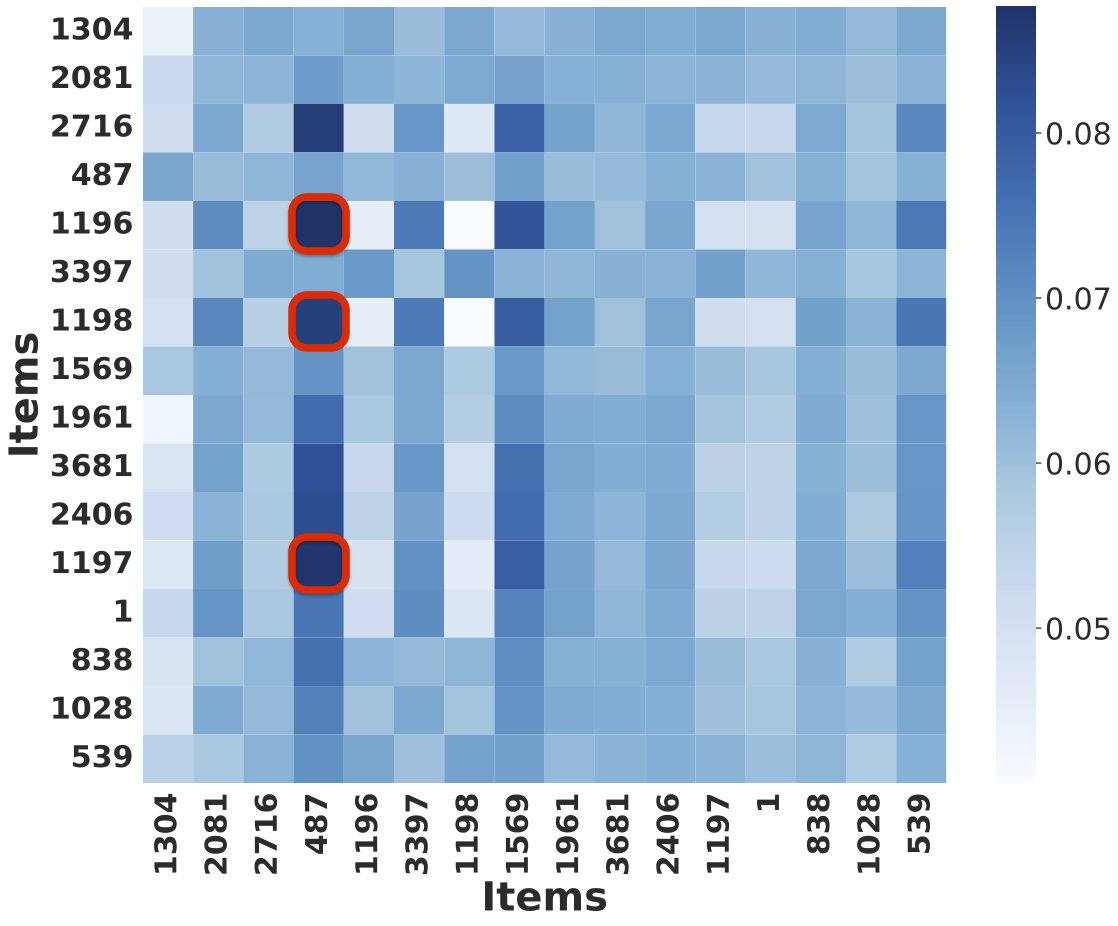}
			\label{fig:itemweight}
            }
             \subfigure[Weights of \ABF (high rating)] {
			\includegraphics[width=0.5\columnwidth]{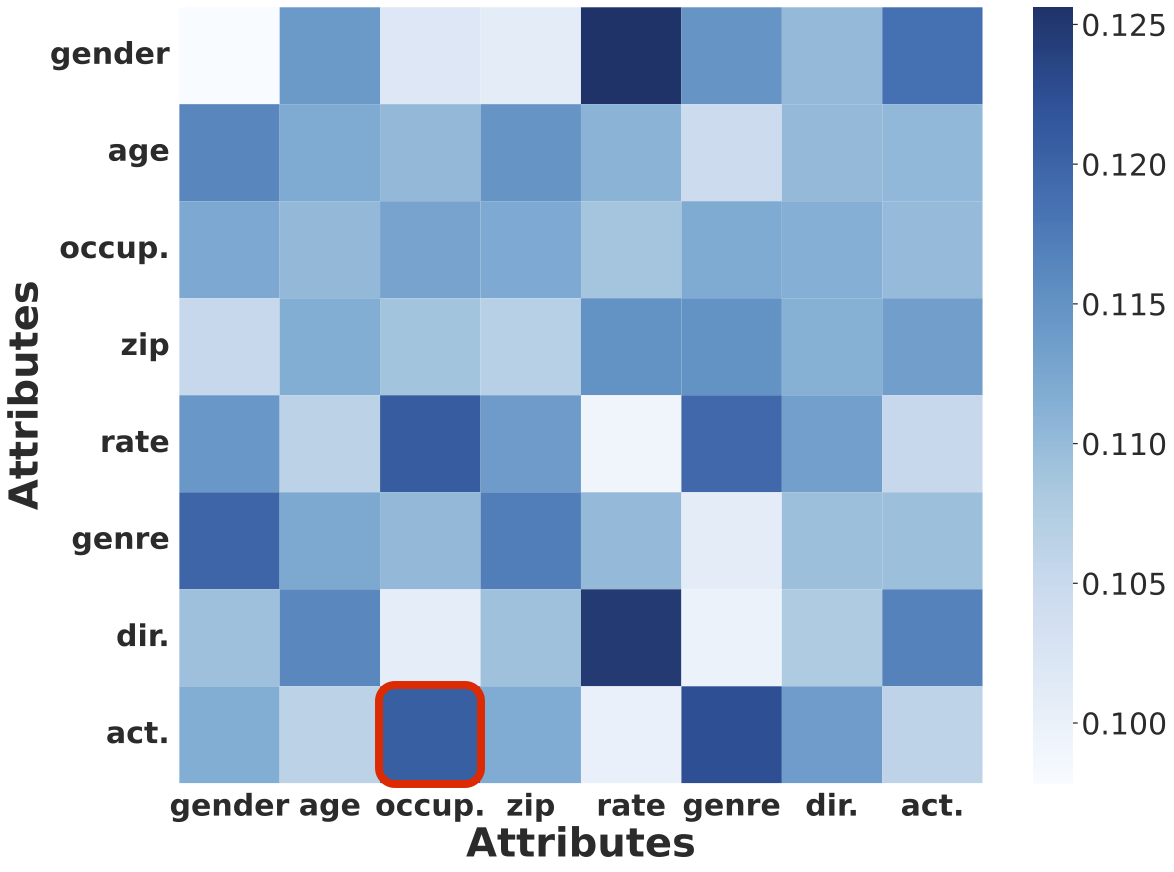}
			\label{fig:featureweight1}
		} 
            \subfigure[Weights of \ABF  (low rating)] {
			\includegraphics[width=0.5\columnwidth]{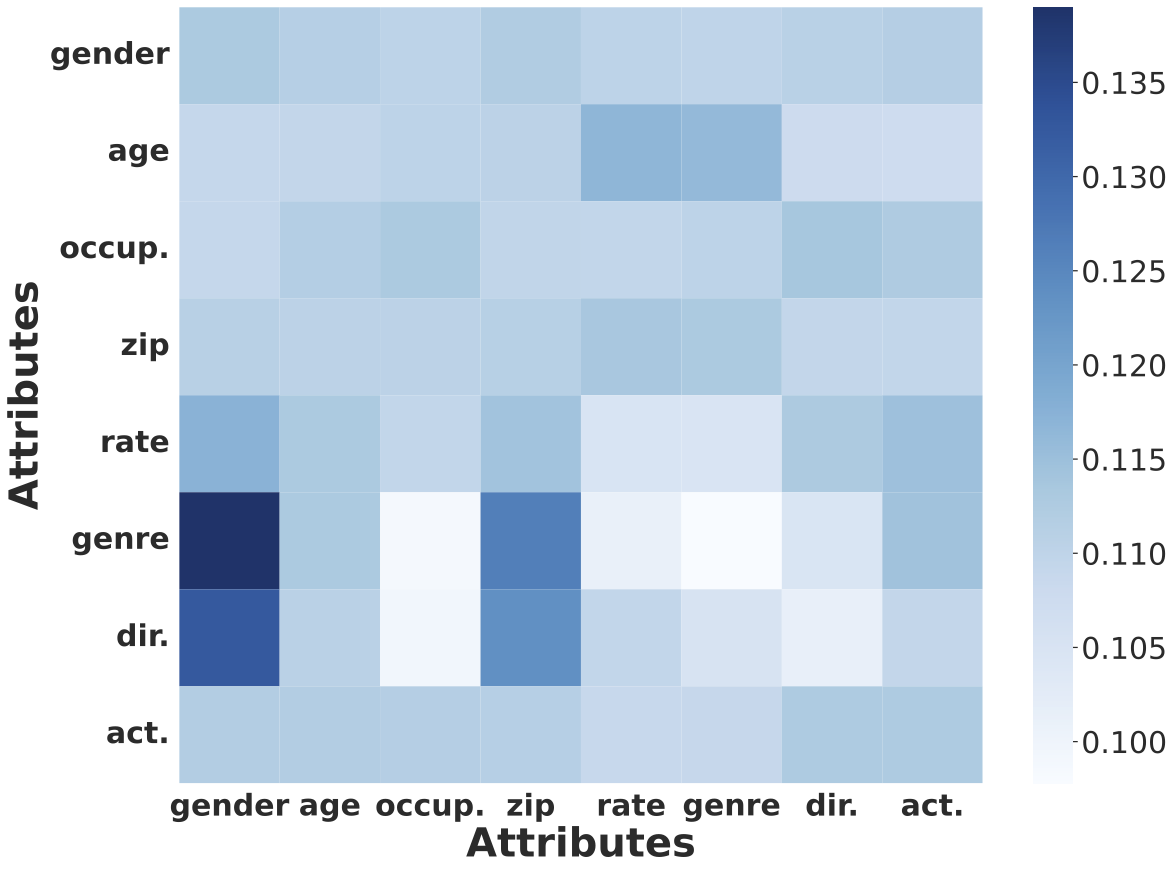}
			\label{fig:featureweight2}
		} \\
	\end{tabular}
	\caption{{Visualizations of the Attention Weights in HIRE}}
        \label{fig:vis}
\end{figure*}

\subsection{Ablation Study}
\label{subsec:ablationstudy}
\etitle{Impacts of the three types of attention layers}.
We conduct an ablation study to investigate the effectiveness of the three types of attention layers, attention between users, items and attributes, in HIRE. Here, we train 6 model variants on \Movielens, where single types or combinations of two types of layers are removed for the model. Table~\ref{tab:layer} lists the test performance of the model variants in 3 cold-start scenarios. The full model with the three types of attention layers achieves the overall best performance, indicating that the heterogeneous interactions from three different perspectives collaboratively contribute to the rating prediction. 
In addition, we find that the attention between items (or attributes) plays a more important role than attention between users in rating prediction, by comparing model variants wo/ Item (or wo/ Attribute) with model variant wo/ User. 
For \Movielens, we speculate that  the similarity between the movies and the similarity between the proprieties of users and movies influences the rating to a larger degree. Meanwhile, another possible reason would be the interactions between users may not be reliable.
That is because there is a counter-intuitive observation on that the model variant wo/ Item (or wo/ Attribute) performs worse than the model variant wo/ User \& Item (or wo/ User \& Attribute). And the model with only attention between users, i.e., wo/ Item \& Attribute, leaves the worst performance among all the model variants. The attention between users may learn irrelevant  interactions, where deploying this layer alone or combining it with another single type of attention layer incurs extra noise for the prediction.
{Furthermore, in our experiments on \Bookcrossing and \Douban datasets, the full model consistently outperforms other variants.}

\comment{
\shadd{
to demonstrate three attention layers are able to model different kinds of interactions,
\shadd{we now present an ablation study on our \CT. We evaluate $6$ \HIM variants and the complete model on \Movielens:} 1) wo/IF: without \ABI and \ABF layers; 2) wo/UF: without \ABU and \ABF; 3) wo/UI:  without \ABU and \ABI; 4) wo/U: without \ABU; 5) wo/I: without \ABI; 6) wo/F: without \ABF. 7) all: complete model, with \ABU, \ABI and \ABF. 
}\shadd{Table~\ref{tab:layer} lists the performance of the \HIM variants on 3 cold-start scenarios on \Movielens. In general, the one with complete \HIM outperforms others. It is because \HIM can learn interactions from 3 perspective, and these heterogeneous interactions can all contribute to the rating prediction.
In addition, the \ABI layer or \ABF layer play more essential roles than \ABU layer as the blocks with \ABF or \ABI layers performs better by a significant margin. The interaction between items or features can truly help predict users' preference.
It is surprisingly to find that when there is only \ABF or \ABI layer in blocks, the model outperforms blocks of wo/I and wo/F. We conjecture that \ABU layer alone or simply combined with one other layer do not work, it must combine with the other two layers which can lead to performance gains. This finding can be explained by the unreliable interaction. Interaction between users tend to be not so reliable thus it works only utilizing additional information i.e. interaction between items and features. 
}
}

\etitle{{Impacts of sampling methods.}}
To study our neighborhood-based sampling strategy for context construction, we compare it with the random sampling strategy and feature similarity sampling strategy as an ablation. For the feature similarity sampling strategy, we compute the cosine similarity of attributes between target users (items) and other users (items). The users (items) with higher scores will be sampled. Fig.~\ref{fig:samplemethod} shows the test result of HIRE on \Movielens, by fixing the number of users and items as $32$. 
Here, our neighborhood-based sampling strategy is better than random sampling in all cases by $1.26\%$ on average. {On the other two datasets, we also achieve similar results, further confirming the effectiveness of the neighborhood-based sampling strategy.}
The reason would be that compared with a fully randomized batch of users/items whose correlations are unknown, neighborhood-based sampling strategy is able to select more relevant neighbor users to construct the prediction context. That promotes HIRE to make more accurate predictions.
Feature similarity sampling strategy tends to perform better than neighborhood-based in the user cold-start scenario. The reason would be that relevant users can be sampled via computing their feature similarity. For cold items, feature similarity is unable to select the most relevant items, which results in poor performance in other two scenarios.

\comment{
\shadd{we compare the effects of batch sampling on \Movielens dataset. We fix the number of users and items as $32$ and compare our 2-hop sampling methods with random sampling. Fig.~\ref{fig:samplemethod} shows the result of 2-hop sampling and random sampling on \Movielens. As we can see, our strategy is better than random sampling in all cases by an average of $1.26\%$. Since 2-hop sampling strategy is able to capture more important neighbors and their interaction information, it is reasonable that 2-hop sampling outperforms random sampling.} 
}

\subsection{Case Study}
\label{subsec:casestudy}

We conduct a case study to investigate whether HIM learns relevant and reliable interactions between users by the 3 layers \ABU, \ABI and \ABF.
We visualize the attention matrices for the prediction on \Movielens in Fig.~\ref{fig:vis}. Here, the darker the cell, the higher the corresponding weight, and the stronger the implicit interaction learned by HIM. 
First, Fig.~\ref{fig:userweight} shows the attention weights in \ABU layer among 16 users for item $1,304$, which is an action and comedy movie, `Butch Cassidy and the Sundance Kid'.
In Fig.~\ref{fig:userweight}, {as the highlighted red rectangle}, user $1,447$ is deeply affected by user $4,926$ and $3,792$. We find that the three users share the same preference on item $1,304$. With the learned user interactions, HIRE predicts the ratings of the three users on item $1,304$ as $3.41$, $4.13$ and $4.20$, which is highly consistent to the ground-truth ratings, $3, 4, 4$, respectively. 
That indicates the interaction between users has a strong correlation to their ratings. 
Second, Fig.~\ref{fig:itemweight} shows the attention weights in \ABI layer among 16 items for user $4,926$,  who is a female student under 18 years old. 
Similarly, we observe that item $487$ is strongly affected by items $1,196$, $1,197$ and $1,198$.
With the learned interaction between items, HIRE predicts the ratings of that user on item $487$ as $3.42$, and on items $1,196$, $1,197$ and $1,198$ as $3.88, 3.95, 3.86$, respectively. That is also consistent to the ground-truth ratings of $3$ for item $487$ and $4$ for the other three items. The results also demonstrate that the learned implicit interactions among items guide the model to predict similar ratings for interacted items. 
Finally, Fig.~\ref{fig:featureweight1} and  \subref{fig:featureweight2} present the attention weights among the 8 attributes in the \ABF layer, for two user-item pairs, user 4,926 and item 1,304, user 1,051 and item 2,081, respectively.  
The former pair has a high rating of $4$.
User $1051$ is a $18-24$ year old male technician/engineer and item $2081$ is a comedy `Trees Lounge'. The user $1051$ gave a low rating $2$ to movie $2,081$. It is reasonable according to the occupation of user and genre of movie.
 As we can see, high rating pair has more attribute interactions. Actor attribute has interactions with occupation in Fig.~\ref{fig:featureweight1}. This means that the actor in the movie may be liked by people in a specific occupation. The latter one has less interactions between movie attributes and user attributes. It may because user has no interest to this movie.
It is worth mentioning that the weight matrices are unsymmetrical due to the computation of attention (Eq.~\eqref{eq:att:weights}),  which indicates the interactions are single direction.

\comment{
In \HIM, we propose three attention layers to model heterogeneous interaction and they consist of multi-head self-attention. We visualize the attention weight in MHSA to show implicit interactions that learned by \HIM. Fig.~\ref{fig:vis} shows visualization of attention maps on \Movielens. The color block represents weight value. The attention weight comes from last \HIM block Head $1$. We take user $4926$, who is a female student under 18 years old and item $1304$, which is an action and comedy movie, called 'Butch Cassidy and the Sundance Kid' as example. 
\shadd{
In \ABU layer, we show attention weights amongst $16$ users for item $1304$ in Fig.~\ref{fig:userweight}. The darker color means higher weight between two users, indicating there exists strong interactions.
User $1447$ is absolutely affected by user $4926$ and $3792$ since they have deep color. We infer that the three user give similar rating to item $1304$. Actually, they give $3$, $4$ and $4$, and HIRE give prediction results $3.4149$, $4.1296$ and $4.1998$, respectively. It validates that these interactions between users learned by model are reliable for rating prediction.}
\shadd{
We show attention weights amongst $16$ items for user $4926$ in Fig.~\ref{fig:itemweight}. As we can see, item $487$ is strongly affected by item $1196$, $1197$ and $1198$. While in reality, user $4926$ rate $4$ to item $487$  , it give score $3$ to item $1196$, $1197$ and $1198$. 
While in our prediction results, the user give $487$, $1196$, $1197$, $1198$ score $3.4213$, $3.8793$, $3.9492$, $3.8644$, respectively.
We conjecture there exists implicit interactions between these items and leads to similar ratings.}
\shadd{Attention weights amongst features for two pairs (user $4926$ and item $1304$, user $1051$ and item $2081$) are shown in Fig.~\ref{fig:featureweight1}, \subref{fig:featureweight2}. 
The former pair has high rating $4$.
User $1051$ is a $18-24$ year old male technician/engineer and item $2081$ is a comedy 'Trees Lounge'. The user $1051$ gave a low rating $2$ to movie $2081$. It is reasonable according to the occupation of user and genre of movie.
 As we can see, high rating pair has more feature interactions. Actor feature have interactions with occupation in Fig.~\ref{fig:featureweight1}. This means that the actor in the movie may be liked by people in a specific occupation. The latter one has less interactions between movie features and user features. It may because user has no interest to this movie.}
It has to be noticed that these interactions are not symmetrical, these directed interactions are reliable in our case study to help rating prediction.
}